\documentclass [12PT]{article}

\usepackage{graphicx}

\usepackage[dvipsnames,usenames]{color}

\definecolor{DarkGreen}{rgb}{0.5,0.8,0.6}   
\definecolor{RGBblack}{rgb}{0.0,0.0,0.0}    

\newcommand{\hei}[1]{\color{black}{#1} \color{black} }
\definecolor{grau}{rgb}{0.8,0.8,0.8}

\newcommand{\chen}[1]{\color{orange}}

\def\hei{\color{black}}

\newcommand{\jj}{\hei \rm}

\usepackage{enumitem}
\usepackage{makecell}
\usepackage{booktabs}
\usepackage{subcaption}
\usepackage{blindtext}
\usepackage{algorithm}
\usepackage{longtable}
\usepackage[table]{xcolor} 
\usepackage{diagbox} 
\usepackage{tikz}
\usetikzlibrary{calc}
\usepackage{colortbl}
\usepackage{multicol}
\usepackage{algorithmic}
\usepackage{natbib}
\usepackage{multirow}
\usepackage{bm}
\usepackage{amsthm,amsmath,amssymb}
\usepackage{mathrsfs}
\usepackage{url}
\usepackage[title]{appendix}
\usepackage{comment}
\usepackage{upgreek}
\usepackage[title]{appendix}
\usetikzlibrary{positioning}
\usetikzlibrary{shapes.geometric, positioning, shadows.blur, backgrounds}

\tikzstyle{block} = [rectangle, rounded corners, thick, minimum width=3cm, minimum height=1.2cm, blur shadow={shadow blur steps=5}, text width=2.8cm, text centered]
\tikzstyle{line} = [thick, -{Triangle[length=6mm, width=3mm]}, line width=1mm, color=black!60]

\newcommand{\cellwidth}{0.7cm}
\newcommand{\cellheight}{0.45cm}

\newcommand{\customcell}[3]{%
  \begin{tikzpicture}
    \node[rectangle, draw=none, minimum width=\cellwidth, minimum height=\cellheight, anchor=south west] (cell) {};
    \ifstrequal{#1}{halfcolor}{
      \fill[yellow!50] ($(cell.north)!0.9!(cell.west)$) rectangle (cell.south east);
      \fill[gray!30] ($(cell.south)!0.1!(cell.east)$) rectangle (cell.north east);
    }{\ifstrequal{#1}{halfcon}{
      \fill[purple!50] ($(cell.north)!0.9!(cell.west)$) rectangle (cell.south east);
      \fill[olive!50] ($(cell.south)!0.1!(cell.east)$) rectangle (cell.north east);
    }
    {\ifstrequal{#1}{halfco}{
      \fill[cyan!50] ($(cell.north)!0.9!(cell.west)$) rectangle (cell.south east);
      \fill[olive!50] ($(cell.south)!0.1!(cell.east)$) rectangle (cell.north east);
    }
        {\ifstrequal{#1}{halfc}{
      \fill[cyan!50] ($(cell.north)!0.9!(cell.west)$) rectangle (cell.south east);
      \fill[purple!50] ($(cell.south)!0.1!(cell.east)$) rectangle (cell.north east);
    }
    {
      \fill[#1] (cell.south west) rectangle (cell.north east);
    }}}}
    \ifstrequal{#2}{}{}{
      \draw[#2, thick] (cell.south west) -- (cell.north east);
    }
    \node[align=center] at (cell.center) {#3};
  \end{tikzpicture}%
}

\usepackage{etoolbox}

\makeatletter

\newcommand{\Rmnum}[1]{\expandafter\@slowromancap\romannumeral #1@}
\makeatother

\usepackage{color}

{
	\title{\bf The Modified Combo i3+3 Design for Novel-Novel Combination Dose-Finding Trials in Oncology}

 	\author{
 		Jiaxin Liu \thanks{Cytel Inc., Shanghai, CHN}, Shijie Yuan\thanks{Department of Statistics and Data Science, The University of Texas at Austin, Austin, USA}, Qiqi Deng\thanks{
Moderna Inc, Cambridge, Massachusetts, USA}, Yuan Ji\thanks{Corresponding email:koaeraser@gmail.com; Department of Public Health Sciences, The University of Chicago, Chicago, USA}
 	}
	\date{\today}
}

\begin{document}
\maketitle

\begin{abstract}
We consider a modified Ci3+3 (MCi3+3) design for  dual-agent dose-finding trials in which both agents are tested on multiple doses. This usually happens when the agents are novel therapies. 
The MCi3+3 design offers a two-stage or three-stage version, depending on the practical need. The first stage begins with single-agent dose escalation,  the second stage launches a  model-free combination dose finding for both agents, and optionally, the third stage follows with a model-based design. 
MCi3+3 aims to maintain a relatively simple framework to facilitate practical application, while also address challenges that are unique to novel-novel combination dose finding. Through simulations, we demonstrate that the MCi3+3 design adeptly manages various toxicity scenarios. It exhibits operational characteristics on par with other combination designs, while offering an enhanced safety profile. The design is motivated and tested for a real-life clinical trial.

\end{abstract}

\textbf{\textit{Keywords}}: Combo Design; Model-Free Designs;  Phase I; Toxicity; Toxicity Probability Interval. 

\newpage
\section{Introduction}
\label{sec:introduction}
\subsection{Overview}
In the evolving landscape of oncology therapeutics, the introduction of combination therapies, particularly with PD-1 inhibitors like pembrolizumab \citep{long2017standard,Baldini2022,BERINSTEIN20193236}, has marked a significant advancement. These therapies often involve administering a novel drug alongside an established one at a fixed dosage, which has been validated for its efficacy and safety. The main aim is to explore the enhanced efficacy of the new drug in combination with a standard treatment regimen. While such trials offer valuable insights, the approach, due to the fixed dosage of the existing drug, mirrors that of single-agent dose-finding studies, thereby simplifying the trial design.

As novel targets and mechanisms continue to emerge like cancer therapeutic vaccines and antibody-drug conjugate, there's an emerging need to simultaneously develop and test multiple new drugs, adding complexity to clinical trials. This shift has led to the exploration of dose combinations, presenting challenges in determining the optimal dosing strategies due to the intricate interplay between the drugs. Different dosing strategies can significantly impact both drugs' efficacy and toxicity profiles. Although some statistical  methods for dual-agent trials are present in the literature, more research is needed to address ongoing issues in combination dose-finding trial.

Hereafter, we use ``combination dose finding", ``dual-agent dose finding", and ``combo dose finding" interchangeably. The traditional Continual Reassessment Method (CRM) \cite{o1990continual} assumes a monotonic relationship between dose levels and toxicity, which may not be valid in dual-agent trials where the toxicity order is not fully known. As a remedy, the POCRM design \cite{wages2011continual} introduces modeling the partial orders under the CRM framework to address the challenge in dual-agent dose finding. In addition, the POCRML design \cite{wages2011dosefinding} extends POCRM by considering a two-stage procedure.  The Waterfall design \citep{ZhangYuan2016} aimied to identify the maximum tolerated dose (MTD) contour by dividing the process into sequential one-dimensional subtrials that leverage previous outcomes to inform subsequent ones. \cite{ManderSweeting2015} introduces a nonparametric dual-agent trial design (PIPE) that simplifies incorporating historical data through conjugate Bayesian inference, ensuring monotonicity in dose-escalation decisions and facilitating varied experimentation along the maximum tolerated contour. In addition, there are methodologies such as the BOIN Combo \citep{LinYin2017}, which adapts the single-agent BOIN design \citep{yuan2015boin} for combination therapies, and the Ci3+3 design \citep{yuan2021ci3+}, an extension of the i3+3 design 
 \citep{liu2020i3+} tailored for dual-agent trials.

While existing methodologies have largely advanced the field for combo dose finding, many challenges and improvements remain. For instance, prior information from single-agent dose finding is not formally modeled, and most methods do not provide sufficiently flexible algorithms that allow thorough and quick exploration of the dose combination space.
Addressing the pressing needs of dual-agent Phase I trial designs, we introduce the Modified Ci3+3 (MCi3+3) design specifically developed  to tackle the issues of dual-agent dose-finding trials.  MCi3+3 is a three-stage design  with the first two stages based on model-free rules and an optional third stage relying on model-based statistical inference.  The first stage begins with   single-agent dose finding using the i3+3 design \citep{liu2020i3+} for each agent in parallel. The trial proceeds to the second stage focusing on exploration of dose combinations (DCs) of the two agents. Importantly, data from the first stage helps determine the starting DCs for the second stage.  
The second stage extends the Ci3+3 design \citep{yuan2021ci3+} with a more intelligent and safer set of rules allowing thorough and efficient exploration of the two-dimensional DC space. Lastly, an optional third stage utilizing logistic regression models can be applied to refine and further explore the space of DCs. The third stage option gives investigators flexibility to go further or stop the trial, which will be demonstrated using numerical examples. 
At the end of the trail, the MCi3+3 select an MTD combination (MTDC) based on all the data in the three stages. 

\subsection{Motivating Example}
The development of the MCi3+3 design is motivated by a real-world example concerning the development of two novel drugs designed to target a previously unexploited molecular pathway in cancer cells. Drug A is a backbone therapy, a small molecule inhibitor that directly interferes with a critical signaling protein within the cancer cells. Drug B, on the other hand, is a supportive monoclonal antibody that targets a surface receptor involved in the same pathway. While Drug A alone has shown promising results in preclinical settings, preclinical data suggests that the addition of Drug B enhances this effect, leading to a synergized effect on tumor reduction. The related first-in-human clinical trial is ready to take place and must determine appropriate dosages for the combination therapy. Since neither drug has been tested in human subjects alone, the trial must consider multiple ascending doses for both drugs and an appropriate dose-escalation design is needed to explore the two-dimensional dose response space. 

Another  challenge is that the two drugs may enter the clinical stage at different times, depending on when the preclinical studies can be completed. Due to the priority of the backbone therapy Drug A, it may be ready for clinical testing earlier than Drug B. If the activity progresses smoothly, both drugs may become available for testing around the same time. Therefore, it is desirable that the dose escalation algorithm can handle either situation. In the former case, it is preferable to start the mono dose escalation of Drug A as soon as it is ready for testing. Early results of the MTD for Drug A may inform the exploration of the dose range space around its MTD, potentially helping to approach the true optimal dose combination faster. Additionally, it is also desirable to add Drug B as soon as it is ready for testing.


This paper is structured as follows: Section \ref{sec:mci3p3_design} introduces the methodology, followed by Section \ref{sec:simulation} presenting the simulation results of the MCi3+3 design. 
Section \ref{sec:trial_sample} demonstrates the major rules of the design using a hypothetical trial sample. Finally, Section \ref{sec:discussion} ends with conclusions and directions for future research.

\section{The MCi3+3 Design }
\label{sec:mci3p3_design}
\subsection{Notation}
For convenience, we use ``drug" and ``agent" interchangeably. Suppose $I$ dose levels of drug A and $J$ levels of drug B are tested. Let $(i,j)$ denote the DC with dose level $i$ for drug A and $j$ for drug B, with the special cases of $(i,0)$ and $(0,j)$ representing dose $i$ of drug A alone and dose $j$ of drug B alone, respectively. Let $x_{ij}=(x_{1i},x_{2j})$ be the actual dosages of DC $(i,j)$, $i$ = 0, $\cdots$, $I$ and $j$ = 0,$\cdots$, $J$. For example, $x_{1i} = 1$ $\upmu$g/kg. 
Assume a binary dose-limiting toxicity (DLT) outcome. Phase I oncology trials enroll patients in cohorts, say a group of three patients per cohort. After a cohort is enrolled and assigned to a dose level for treatment, the patients are followed for three to four weeks to evaluate drug safety and record any DLT outcomes. Let $n_{ij}$ denote the number of patients assigned to DC $(i,j)$ and $y_{ij}$ the number of patients out of $n_{ij}$ experienced DLT.  
Let $p_{ij}$ denotes the DLT probability of DC $(i, j)$, we assume
$$y_{ij} \mid n_{ij},p_{ij} \sim Bin(n_{ij},p_{ij}).$$

In the MCi3+3 design, we consider the up-and-down decision rules and utilize an equivalence interval (EI) for dose escalation. Let $p_T$ be the target toxicity probability of a maximum tolerated DC (MTDC) and assume the EI, taking the form of $[p_T - \epsilon_1, p_T + \epsilon_2]$, defines an interval so that a DC with toxicity probability falling into the EI is considered an acceptable MTDC. The first and second stages of MCi3+3 are anchored based on the rules in the i3+3 design, summarized in Table \ref{tbl:i3+3_algorithm}. 
Decisions $E$, $S$, and $D$ represent escalation to the next higher dose, stay at the current dose, and de-escalation to the next lower dose, respectively. 

 \begin{table}[!htbp]
       \begin{center}
        \caption{The  decision rules in the i3+3 design for a single-agent dose finding. Notation: $d$ represents the current dose being investigated in the trial; $n_d$ and $y_d$ denote the number of patients enrolled and those with DLT at dose $d$, respectively. }
        \label{tbl:i3+3_algorithm}
	\begin{tabular}{|p{8cm} |{c}| c|}
		\hline
		{\it Condition} & {\it Decision} & {\it Dose for next cohort}\\
		\hline 
		$\frac{y_d}{n_d}$ below EI	& Escalation ($E$)  & $d+1$ \\ \hline
		$\frac{y_d}{n_d}$ inside EI	& Stay ($S$) & $d$ \\ \hline
		$\frac{y_d}{n_d}$ above EI and $\frac{y_d-1}{n_d}$ below EI	& Stay ($S$) & $d$ \\ \hline
		$\frac{y_d}{n_d}$ above EI and $\frac{y_d-1}{n_d}$ inside EI	& De-escalation ($D$) & $d-1$\\ \hline
		$\frac{y_d}{n_d}$ above EI and $\frac{y_d-1}{n_d}$ above EI	& De-escalation ($D$)  & $d-1$\\
		\hline
	\end{tabular}
 \end{center}
\end{table}


The MCi3+3 design comprises three stages with Stage III optional. We present the three stages next.

\subsection{Stage \Rmnum{1} -  Single-Agent Dose Finding} 
\label{sec:single_agent}
Motivated by the trial example, we assume both agents are tested for the first time in humans. Therefore, the first stage of the MCi3+3 design is related to single-agent dose finding for each agent. 
The objective of this stage is to explore the toxicity profiles of various doses of each agent when it is administered alone, and therefore inform  an appropriate starting DC based on single-agent data. We consider using the i3+3 design \citep{liu2020i3+} for this stage following the algorithm below. The algorithm is applied 
to each agent independently as there are no DCs in this stage.  
\begin{enumerate}
    \item Enroll the initial two cohorts, one for each drug,  at their   starting doses, $(1,0)$ and $(0,1)$.
    \item After each cohort completes the DLT evaluation period, use the decisions of the i3+3 design in Table \ref{tbl:i3+3_algorithm} to determine escalation ($E$) to the next higher dose, de-escalation ($D$) to the next lower dose, or staying at the current dose ($S$). 
    \item Enroll a new cohort of patients for each drug at the corresponding dose in step 2. 
    \item Repeat steps 2-3 until for the first time, a dose is assigned a decision $S$ or $D$ for each drug.  Denote the two doses $(i_0 + 1, 0)$ and $(0, j_0+1)$ for both drugs. If all the doses, including the highest dose of each drug, are assigned decision $E$, then $(i_0+1)= (I+1)$ and $(j_0+1)=(J+1).$
    \item Enter the second stage for DC finding. The starting DCs are 
    $(i_0,1)$ and $(1,j_0)$, if $i_0, j_0 \ge 1.$ Otherwise, there is only one starting DC for the second stage, DC $(1,1).$  
\end{enumerate}

The algorithm applies the i3+3 rules for each agent until for the first time the decision is not escalation ($E$). The starting DC for the subsequent combination dose finding will be determined to be the next lower dose of that agent and the first (lowest dose) of the other agent. If no doses for either agent receive an $E$, the starting DC will be DC (1,1), to be safe.




\subsection{Stage \Rmnum{2} - Rule-Based DC Finding}
\label{sec:stage2}
\paragraph{DC-Finding Algorithm} In the second stage, the trial shifts to DC finding in which up to two DCs can be tested at a time. Let's denote these DCs the ``current DCs''. We will use a generalized up-and-down decision rule (GUD) to determine the DCs based on the observed data from the current DCs. 
The GUD is developed to accommodate the possibility of having multiple candidate DCs for each decision of $E$, $S$, or $D$. For example, a decision $E$ at DC $(i,j)$ means escalation to a higher DC, which could be $(i+1,j)$ or $(i,j+1)$ as candidate DCs for the next cohort. To resolve the issue and select up to two candidate DCs for the next cohorts, we extend the rules in the Ci3+3 design \citep{yuan2021ci3+}.

First, we do not allow simultaneous change of both doses of a DC when escalation. 
For example, escalation from $(i,j)$ to $(i+1, j+1)$ is not allowed. Second, when multiple candidate DCs are available based on the decisions of the current DCs, we select up to two DCs with the highest utility, which will be defined later. 

We present Stage II of the MCi3+3 design next. For convenience, denote $\sigma = \{ (i,j),0\le i \le I, 0 \le j \le J\} \backslash (0,0)$, i.e., the set of all $(I+1)*(J+1)$ DCs excluding $(0,0)$. Denote ${\bm n} = (n_{ij}, (i,j)\in \sigma)$ and ${\bm y} = (y_{ij}, (i,j) \in \sigma).$
\begin{enumerate}
    \item The starting DCs are $\{(i_0, 1), (1, j_0)\}$ or $(1, 1)$ according to stage I step 5. 
    \item MCi3+3 allows up to two current DCs to be tested at each step. Consider a current DC  $(i, j)$. 
      Calculate the dosing decision $E$, $S$, or $D$ for DC $(i, j)$ based on data $(n_{ij}, y_{ij})$ and the rules of the i3+3 design in Table \ref{tbl:i3+3_algorithm}. Denote the corresponding decision $S_{ij}$, where $S_{ij} \in \{D, S, E\}.$
      \item {\bf [Add DCs to $\mathbf{\Omega}$]} Let $\Omega$ denote the candidate DCs for the next cohorts. 
    \begin{itemize}
        \item[a.] If $S_{ij} = E$, i.e., the i3+3 decision of DC $(i, j)$ is $E$, DCs $(i+1, j)$ and $(i, j+1)$ are added to $\Omega$.
        \item[b.] Else if $S_{ij} = S$, i.e., the i3+3 decision of DC $(i, j)$ is $S$, 
        \begin{itemize}
        \item Add DCs $(i,j)$, $(i+1, j-1)$, and $(i-1, j+1)$ to $\Omega;$
         \item If $n_{i+1,j-1} \neq 0$, $n_{i+2,j-2}=0$, and $S_{i+1, j-1} = E$ or $S$,  DC $(i+2, j-2)$ is added to $\Omega$. 
        \item If $n_{i-1,j+1} \neq 0$, $n_{i-2,j+2}=0$, and $S_{i-1, j+1} = E$ or $S$, DC $(i-2,j+2)$ is added to $\Omega$. 
        \end{itemize}
        \item[c.] Else if $S_{ij} = D$, i.e., the i3+3 decision at DC $(i, j)$ is $D$, DCs $(i-1,j)$ and $(i,j-1)$ are added to $\Omega$.
        \item[d.] A DC $(i, j)$ is higher than ($i^{'},j^{'})$ if
        \begin{itemize}
            \item $i \textgreater i^{'}$ and $j \textgreater j^{'} $,
            \item or $i \textgreater i^{'}$  and $j = j^{'} $,
            \item or  $i = i^{'}$ and $j \textgreater j^{'} $.
        \end{itemize}
        \item[e.] Similarly, a DC $(i, j)$ is lower than ($i^{'},j^{'})$ if
        \begin{itemize}
            \item $i \textless i^{'}$ and $j \textless j^{'} $,
            \item or $i \textless i^{'}$  and $j = j^{'} $,
            \item  or $i = i^{'}$ and $j \textless j^{'} $.
        \end{itemize}
        \end{itemize}
        \item {\bf [Prune DCs in $\mathbf{\Omega}$]} Denote the current available trial data $(\bm{n}, \bm{y}) $ at all the DCs with $n_{ij} \neq 0$. Compute the dosing decisions on all the these DCs based on the i3+3 design (Table \ref{tbl:i3+3_algorithm}).
        \begin{itemize}
           \item[a.] For any DC $(i,j)$, if $S_{ij} = E$ and $n_{ij} > 0$, all the DCs lower than the DC $(i,j)$, denoted as set $\sigma_E(\bm{n}, \bm{y})$, are removed from $\Omega$.
            \item[b.] For any DC $(i,j)$, if $S_{ij} = D$ and $n_{ij} > 0$, all the DCs higher than the DC $(i,j)$, denoted as set  $\sigma_D(\bm{n}, \bm{y})$, are removed from $\Omega$. 

        \end{itemize}
    \item  {\bf [Finalize DCs in $\mathbf{\Omega}$]}	Denote the current DCs $A_k$, $k=1,$ or $2$. Denote $\Omega$ the final set of candidate DCs from Steps 3-4 above. 
    \begin{itemize}
        \item [a.] If $\Omega \neq \emptyset$, then check if any current DC $A_k \in \Omega$. If there is an $A_k \in \Omega$ but its decision $S_{A_k}$ is not $S$, remove $A_k$ from $\Omega$.
        \item[b.] 	If $\Omega = \emptyset$, let $\Omega = (\sigma - \sigma_D) \cap (\sigma - \sigma_E)$, the set of admissible DCs.
    \end{itemize}
    \item  {\bf [Select DCs for the next cohort]}	Select up to two DCs from $\Omega$  with the highest utilities $U_{ij}$ and treat the next cohorts at the two DCs. If there are more than two DCs with the same highest utility, select two randomly. The utility $U_{ij}$ is computed according to the following steps. 
    \begin{itemize}
        \item[a.] Compute $U'_{ij}=\mbox{Pr}\{p_{ij} \in \mbox{EI}  \mid   ( n_{ij},y_{ij})\}$ with $Beta(0.05,0.05)$ as the prior of $p_{ij}$. 
        \item[b.] Let $\delta_{ij}=(x_{1i} + x_{2j})*\epsilon,$ where $\epsilon$ is a tiny fraction, say $\epsilon=10^{-6}.$
        \item[c.] If $y_{ij}/n_{ij} \le p_T,$ the utility of DC $(i,j)$ is given by $U_{ij} = U'_{ij} + \delta_{ij}$. Else if $y_{ij}/n_{ij} > p_T,$ the utility of DC $(i,j)$ is given by $U_{ij} = U'_{ij} - \delta_{ij}$. This rule is used to break ties when two DCs have the same $U'$ values. 
    \end{itemize}   
\end{enumerate}

\begin{figure}[!ht]
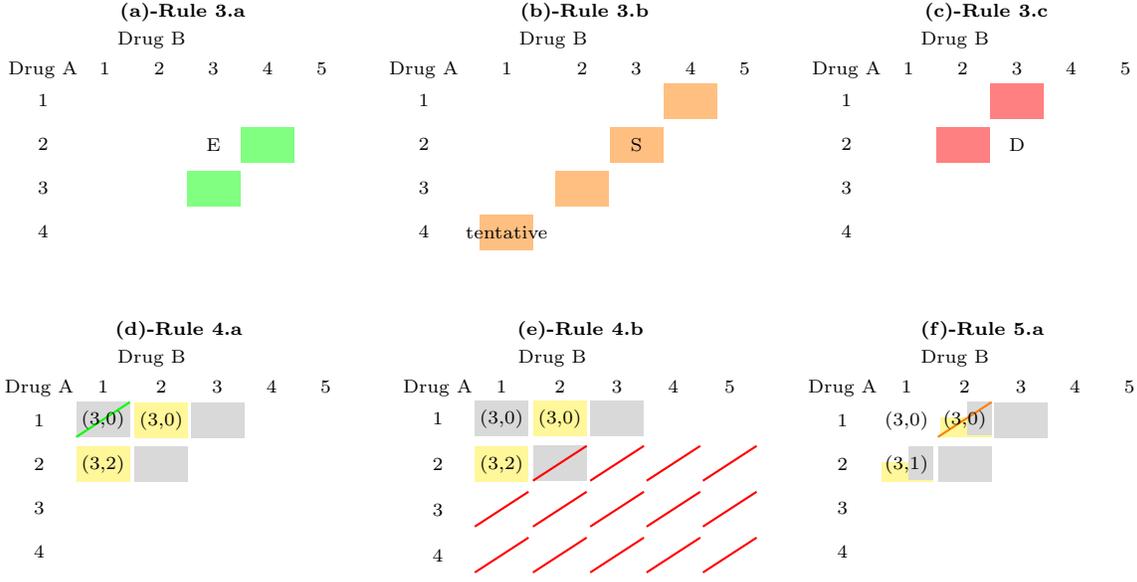

\centering
\setlength{\tabcolsep}{0.1pt}
\renewcommand{\arraystretch}{0.8}

\fontsize{7pt}{8pt}\selectfont

\begin{tabular}{ccc}

\minipage[t]{0.3\textwidth}
\centering
\textbf{ (a)-Rule 3.a}\\
\begin{tabular}{cccccc}
\multicolumn{5}{c}{Drug B} \\
Drug A & 1 & 2 & 3 & 4 & 5 \\
\raisebox{1.5ex}[0pt][0pt]{1} & \customcell{white}{}{} & \customcell{white}{}{} & \customcell{white}{}{} & \customcell{white}{}{} & \customcell{white}{}{} \\
\raisebox{1.5ex}[0pt][0pt]{2} & \customcell{white}{}{} & \customcell{white}{}{} & \customcell{white}{}{E} & \customcell{green!50}{}{} & \customcell{white}{}{} \\
\raisebox{1.5ex}[0pt][0pt]{3} & \customcell{white}{}{}  & \customcell{white}{}{} & \customcell{green!50}{}{} & \customcell{white}{}{} & \customcell{white}{}{} \\
\raisebox{1.5ex}[0pt][0pt]{4} &\customcell{white}{}{} &  \customcell{white}{}{}&  \customcell{white}{}{}& \customcell{white}{}{}& \customcell{white}{}{}\\
\end{tabular}
\endminipage
&

% ...
\minipage[t]{0.3\textwidth}
\centering
\textbf{ (b)-Rule 3.b}\\
\begin{tabular}{cccccc}
\multicolumn{5}{c}{Drug B} \\
Drug A & 1 & 2 & 3 & 4 & 5 \\
\raisebox{1.5ex}[0pt][0pt]{1} & \customcell{white}{}{} & \customcell{white}{}{} & \customcell{white}{}{} & \customcell{orange!50}{}{} & \customcell{white}{}{} \\
\raisebox{1.5ex}[0pt][0pt]{2} & \customcell{white}{}{} & \customcell{white}{}{} & \customcell{orange!50}{}{S} & \customcell{white}{}{} & \customcell{white}{}{} \\
\raisebox{1.5ex}[0pt][0pt]{3} & \customcell{white}{}{}  & \customcell{orange!50}{}{} & \customcell{white}{}{} & \customcell{white}{}{} & \customcell{white}{}{} \\
\raisebox{1.5ex}[0pt][0pt]{4} &\customcell{orange!50}{}{tentative} &  \customcell{white}{}{}&  \customcell{white}{}{}& \customcell{white}{}{}& \customcell{white}{}{}\\
\end{tabular}
\endminipage
&

\minipage[t]{0.3\textwidth}
\centering
\textbf{ (c)-Rule 3.c}\\
\begin{tabular}{cccccc}
\multicolumn{5}{c}{Drug B} \\
Drug A & 1 & 2 & 3 & 4 & 5 \\
\raisebox{1.5ex}[0pt][0pt]{1} & \customcell{white}{}{} & \customcell{white}{}{} & \customcell{red!50}{}{} & \customcell{white}{}{} & \customcell{white}{}{} \\
\raisebox{1.5ex}[0pt][0pt]{2} & \customcell{white}{}{} & \customcell{red!50}{}{} & \customcell{white}{}{D} & \customcell{white}{}{} & \customcell{white}{}{} \\
\raisebox{1.5ex}[0pt][0pt]{3} & \customcell{white}{}{}  & \customcell{white}{}{} & \customcell{white}{}{} & \customcell{white}{}{} & \customcell{white}{}{} \\
\raisebox{1.5ex}[0pt][0pt]{4} &\customcell{white}{}{} &  \customcell{white}{}{}&  \customcell{white}{}{}& \customcell{white}{}{}& \customcell{white}{}{}\\
\end{tabular}
\endminipage
\end{tabular}

\vspace{20pt}

\begin{tabular}{ccc}

\minipage[t]{0.3\textwidth}
\centering
\textbf{(d)-Rule 4.a} \\
\begin{tabular}{cccccc}
\multicolumn{5}{c}{Drug B} \\
Drug A & 1 & 2 & 3 & 4 & 5 \\
\raisebox{1.5ex}[0pt][0pt]{1} & \customcell{gray!30}{green}{(3,0)} & \customcell{yellow!50}{}{(3,0)} & \customcell{gray!30}{}{} & \customcell{white}{}{} & \customcell{white}{}{} \\
\raisebox{1.5ex}[0pt][0pt]{2} & \customcell{yellow!50}{}{(3,2)} & \customcell{gray!30}{}{} & \customcell{white}{}{} & \customcell{white}{}{} & \customcell{white}{}{} \\
\raisebox{1.5ex}[0pt][0pt]{3} & \customcell{white}{}{}  & \customcell{white}{}{} & \customcell{white}{}{} & \customcell{white}{}{} & \customcell{white}{}{} \\
\raisebox{1.5ex}[0pt][0pt]{4} &\customcell{white}{}{} &  \customcell{white}{}{}&  \customcell{white}{}{}& \customcell{white}{}{}& \customcell{white}{}{}\\
\end{tabular}
\endminipage
&

% ...
\minipage[t]{0.3\textwidth}
\centering
\textbf{(e)-Rule 4.b} \\
\begin{tabular}{cccccc}
\multicolumn{5}{c}{Drug B} \\
Drug A & 1 & 2 & 3 & 4 & 5 \\
\raisebox{1.5ex}[0pt][0pt]{1} & \customcell{gray!30}{}{(3,0)} & \customcell{yellow!50}{}{(3,0)} & \customcell{gray!30}{}{} & \customcell{white}{}{} & \customcell{white}{}{} \\
\raisebox{1.5ex}[0pt][0pt]{2} & \customcell{yellow!50}{}{(3,2)} & \customcell{gray!30}{red}{} & \customcell{white}{red}{} & \customcell{white}{red}{} & \customcell{white}{red}{} \\
\raisebox{1.5ex}[0pt][0pt]{3} & \customcell{white}{red}{}  & \customcell{white}{red}{} & \customcell{white}{red}{} & \customcell{white}{red}{} & \customcell{white}{red}{} \\
\raisebox{1.5ex}[0pt][0pt]{4} &\customcell{white}{red}{} &  \customcell{white}{red}{}&  \customcell{white}{red}{}& \customcell{white}{red}{}& \customcell{white}{red}{}\\
\end{tabular}
\endminipage
&

\minipage[t]{0.3\textwidth}
\centering
\textbf{(f)-Rule 5.a} \\
\begin{tabular}{cccccc}
\multicolumn{5}{c}{Drug B} \\
Drug A & 1 & 2 & 3 & 4 & 5 \\
\raisebox{1.5ex}[0pt][0pt]{1} & \customcell{white}{}{(3,0)} & \customcell{halfcolor}{orange}{(3,0)} & \customcell{gray!30}{}{} & \customcell{white}{}{} & \customcell{white}{}{} \\
\raisebox{1.5ex}[0pt][0pt]{2} & \customcell{halfcolor}{}{(3,1)} & \customcell{gray!30}{}{} & \customcell{white}{}{} & \customcell{white}{}{} & \customcell{white}{}{} \\
\raisebox{1.5ex}[0pt][0pt]{3} & \customcell{white}{}{}  & \customcell{white}{}{} & \customcell{white}{}{} & \customcell{white}{}{} & \customcell{white}{}{} \\
\raisebox{1.5ex}[0pt][0pt]{4} &\customcell{white}{}{} &  \customcell{white}{}{}&  \customcell{white}{}{}& \customcell{white}{}{}& \customcell{white}{}{}\\
\end{tabular}
\endminipage
\end{tabular}

\caption{
 (a)-(c): Green, orange, and red boxes correspond to candidate DCs for decisions $E$, $S$, and $D$, respectively. (d)-(e): DCs (1,2) and (2,1) are the current DCs marked by yellow boxes.  Grey boxes are the candidate DCs. All the DCs with a $\slash$ sign (red or green) cannot be a candidate DC due to being either too risky (red) or too conservative (green); 
therefore only DC (1,3) is allowed for further testing.  (f): DCs (1,2) and (2,1) are  the current DCs marked by yellow, and DCs (1,2), (1,3), (2,1) and (2,2) are initially considered as candidate DCs for further testing, marked by grey. However,  Rule 5.a eliminates DC (1,2) and Rule 6 selects two DCs (2,1) and (2,2) as the final candidates for further testing.}
\label{fig:additional_rule}
\end{figure}

Figure \ref{fig:additional_rule} illustrates the above Stage II rules with graphical displays. Each sub-figure is used to illustrate a specific rule described as the sub-figure title. Specifically,
later in Figure \ref{fig:trial_sample}  Section \ref{sec:trial_sample}, we provide a stylized example to fully illustrate these rules using a trial example. 

\paragraph{Continuous Monitoring} We consider a change-point model to continuously monitor the progress of the trial. According to the change-point model when $\mbox{Pr}\{p_{ij} \in \mbox{EI} \mid (\bm{n}, \bm{y})\} > \eta$, 
for a moderate value $\eta$ say 0.4, there is reasonable evidence that at least one DC $(i,j)$ ($i>0, j>0$) is within the target range for the MTDC. At this point, an optional stage III may start. The change-point model is given by 
\begin{equation}
\begin{aligned}
    \mbox{logit}(p_{ij}) = \beta_0 +(\beta_1 x_{1i} +\beta_2 x_{2j}+\beta_3 x_{1i} x_{2j})I(x_{1i} \le x_{1max})I(x_{2j} \le x_{2max})+(\beta_4 x_{1i} \\+\beta_5 x_{2j}+\beta_6 x_{1i} x_{2j})(1-I(x_{1i} \le x_{1max})I(x_{2j} \le x_{2max})) \label{eq:changepoint}
\end{aligned}
\end{equation}
where $x_{1max}$ and $x_{2max}$ are the maximum dose levels tested so far for drugs A and B, respectively. 
The priors for $\beta_1$ to $\beta_3$ are set as log-normal distributions, characterized by a mean of $-2$ and a large variance. We utilize a normal prior with a mean of $-4$ for $\beta_0$ to indicate the toxicity level when no dose combinations  are administered. 
To accommodate situations involving untried DCs, we specify a normal prior for $\beta_4$ to $\beta_6$ with a mean of $-2$ and a notably larger variance of 50, in contrast to the smaller variances assigned to parameters $\beta_1$ through $\beta_3$. Essentially, the change-point model aims to utilize the data from explored DCs only to inform the dose-toxicity relationship on the explored DCs. This avoids the potential lack of model fitting due to a large proportion of unexplored doses in the early stage of the trial.

\subsection{Stage \Rmnum{3} -  Model-Based DC Finding}
In the optional stage \Rmnum{3}, we consider a fully model-based inference since sufficient data have been accumulated across many DCs in the previous two stages. The model-based inference allows all the previous data to be used for decision making. We consider a different logistic model dropping change point in \eqref{eq:changepoint} because we want to extrapolate to DCs beyond $x_{1max}$ and $x_{2max}.$ Specifically, 
\begin{equation}
    \mbox{logit}(p_{ij}) = \beta_0 + \beta_1 x_{1i} +\beta_2 x_{2j} + \beta_3 x_{1i} x_{2j}, \label{eq:stage3}
\end{equation}
where the priors for $\beta$'s are the same as for \eqref{eq:changepoint}. At each step of Stage III,  the DC with the highest utility $U_{ij}=\mbox{Pr}\{p_{ij} \in \mbox{EI}  \mid  (\bm{n}, \bm{y})\}$ in the admissible set is considered the candidate DC for further testing. 

\subsection{Safety Rules and MTDC Selection} 
\label{sec:safety_rule}
We consider two safety rules (SRs) to enhance the protection of patients. \\
 {\bf SR 1}:  Following the i3+3 design \citep{liu2020i3+}, apart from $E$, $S$ and $D$, DU is also defined by calculating the posterior probability $\mbox{Pr}\{p_{ij} > p_T \mid (n_{ij},y_{ij})\} > \eta$
 and the threshold  $\eta$  is close to 1, say 0.95. This probability is calculated based on the prior $Beta(\alpha_0, \beta_0),$ $\alpha_0=\beta_0=0.05$ for 
 $p_{ij}$. If the condition is met and $n_{ij} \ge 3$, then the current and all higher DCs
 are removed from the trial.
If the lowest dose is removed from the trial, then the trial is stopped. \\
{\bf SR 2}: Define $(\sigma - \sigma_D) \cap (\sigma - \sigma_E)$ a set of admissible DCs. If it is empty at any point of the trial, the trial is stopped.

\paragraph{MTDC Selection}

At the end of the study, if the trial stops early due to either of the two safety rule, no DC is selected as the MTDC.  
Otherwise, an isotonic regression is applied to calculate the posterior means of the DCs $\{(i,j),  i>0, j>0\}$. The posterior mean for DC $(i,j)$ is estimated to be $(a_0 + y_{ij})/(a_0 + b_0 + n_{ij})$, with $a_0=b_0=0.005.$ 
Let $\hat{p}_{ij}$ be the isotonic-transformed posterior mean for DC $(i,j)$. The DC with $\hat{p}_{ij}$ the closest  to the target toxicity rate $p_T$ is selected as the MTDC, i.e., 
\begin{align}
    \mathop{\arg\max}\limits_{(i,j)} | \hat{p}_{ij} - p_T |.
\end{align}
Untested DCs (i.e., $n_{ij} =0$) or DCs that are eliminated due to SR 1 are not eligible for MTDC consideration. 

\section{Simulation}
\label{sec:simulation}

We conduct a comparative simulation of the  MCi3+3, Ci3+3 and POCRML designs to assess their operating characteristics (OCs). We consider seven scenarios based on those proposed in the Ci3+3 design \citep{yuan2021ci3+} featuring multiple true MTDCs. 
See Table \ref{tbl:true_toxicity} for scenario details. Since the Ci3+3 design has already been compared to other popular rule-based methods and demonstrated comparable results, we focus the comparison to the Ci3+3 design here rather than including many other designs. Apart from Ci3+3 design, we also compare with POCRML design as it is one of the best performing methods for dual agents.

We consider two settings of POCRML (Appendix \ref{sec:partial_order}), each defined by the simple order choice of the DCs under the partial order.  In the first setting, we assume there are five possible simple orders and assigning each simple order $m$ a probability of $p(m)=1/5$, $m \in \{1,2,3,4,5\}$. This version is labeled ``POCRML-5". In the second setting, we use the true simple order of the scenario for the POCRML design, giving it the optimal performance. We label this version of POCRML ``POCRML-1". 
The two settings are introduced in detail in Appendix \ref{sec:partial_order}.
We implement both versions of POCRML using the R package ``pocrm".  

\begin{figure}[!htbp]
    \centering
    \includegraphics[width=0.85\textwidth]{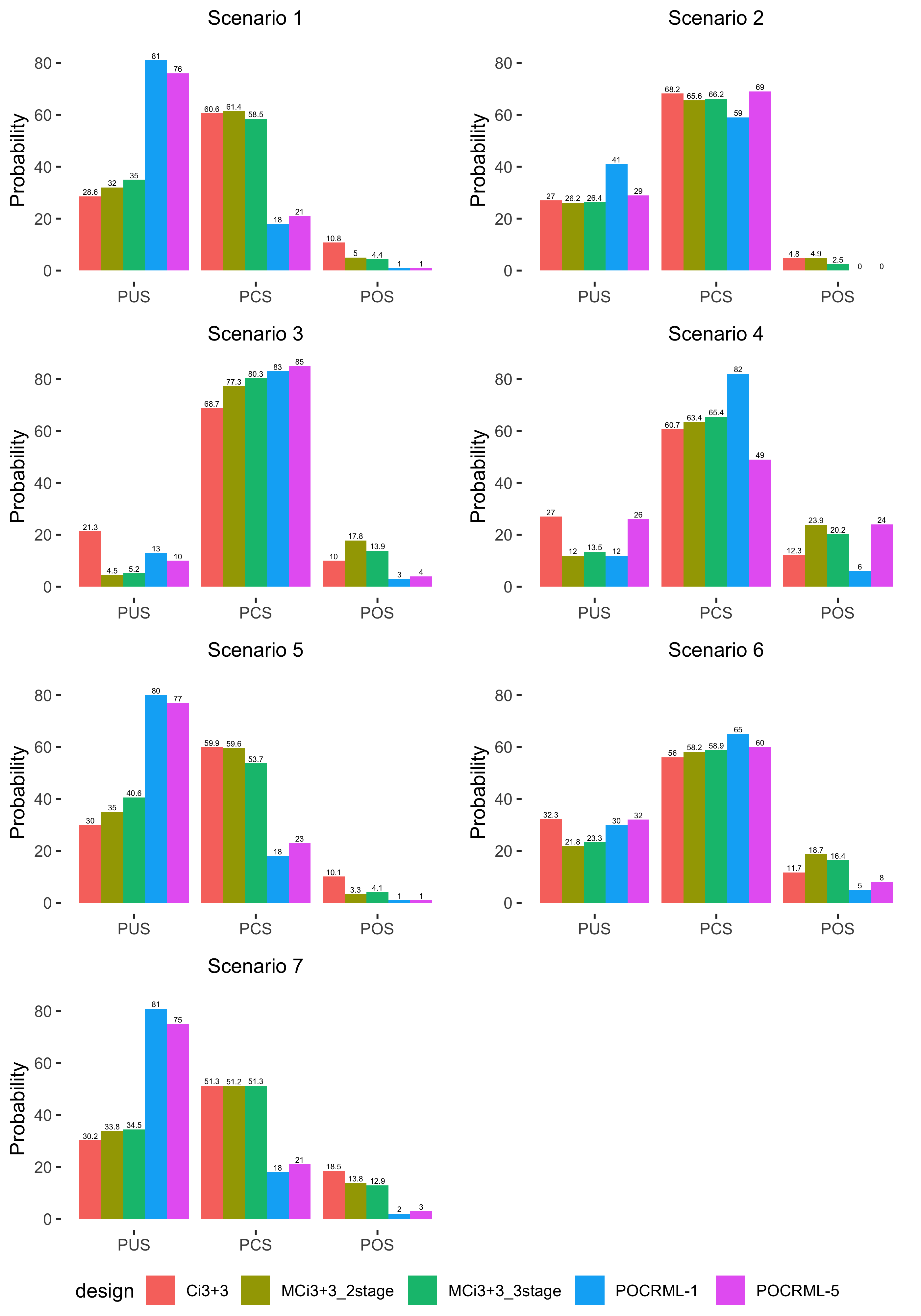}
    \caption{MTDC Selection of the Designs.} 
    \label{fig:mtd_selection}
\end{figure}

\begin{figure}[!htbp]
    \centering
    \includegraphics[width=0.85\textwidth]{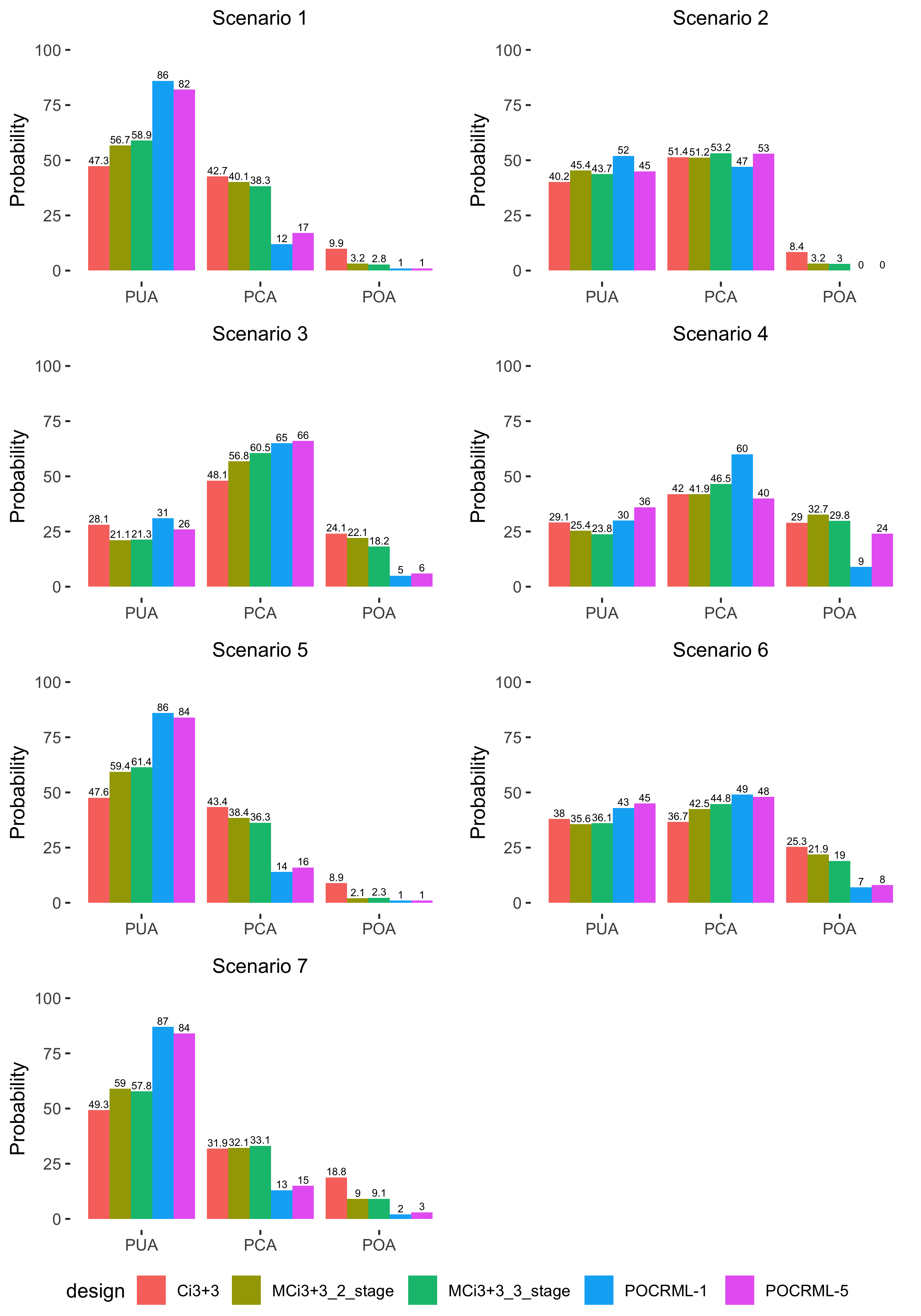}
    \caption{Patient Allocation of the Designs.} 
    \label{fig:patient_allocation}
\end{figure}

We set the maximum sample size as 96 for MCi3+3 and target $p_T$ =0.3. We reduce the maximum sample size to 74 for the Ci3+3 and POCRML designs accordingly since they do not include single-agent dose finding. Also, we let
Ci3+3 starts from DC $(1,1).$ For each design, $1,000$ trials are simulated and their OCs are summarized in Figures \ref{fig:mtd_selection} and \ref{fig:patient_allocation}.


\newcommand{\scenario}[2]{ 
  \begin{minipage}[t]{0.45\textwidth}
    \centering
    \textbf{Scenario #1} \\[0.01ex]
    $\begin{array}{c|cccccc}
    \multicolumn{1}{c}{} & \multicolumn{6}{c}{\text{Drug B}} \\
    \cline{2-7}
    \text{Drug A} & 0 & 1 & 2 & 3 & 4 & 5 \\
    \hline
     #2 \\
    \end{array}$
    \vspace{1ex}
  \end{minipage}
}

\begin{table}[!htbp]
\caption{True toxicity probabilities of DCs in seven scenarios. The bold font indicates a true MTDC. Multiple true MTDCs might be present in a scenario as long as their toxicity probabilities fall into the EI. Here, $p_T=0.3$ and EI=$(0.25, 0.35).$}
\centering
\small
\scenario{1}{
  0& & 0.02 & 0.04 & 0.06 & 0.08 & 0.09 \\
  1&0.02 & 0.04 & 0.08 & 0.12 & 0.16 & 0.18 \\
  2&0.05 & 0.10 & 0.14 & 0.18 & 0.22 & \textbf{0.26} \\
  3&0.08 & 0.16 & 0.20 & 0.24 & \textbf{0.28} & \textbf{0.30} \\
  4&0.11 & 0.22 & \textbf{0.26} & \textbf{0.30} & \textbf{0.34} & 0.36
}
\hspace{6ex} 
\scenario{2}{
  0&& 0.04 & 0.09 & 0.14 & 0.145 & 0.15 \\
  1&0.04 & 0.08 & 0.18 & \textbf{0.28} & \textbf{0.29} & \textbf{0.3} \\
  2&0.045 & 0.09 & 0.19 & \textbf{0.29} & \textbf{0.3} & \textbf{0.32} \\
  3&0.05 & 0.1 & 0.2 & \textbf{0.3} & \textbf{0.31}& \textbf{0.35}\\
  4&0.055 & 0.11 & 0.21 & \textbf{0.31} & 0.41 & 0.51
}
\scenario{3}{
  0&& 0.02 & 0.045 & 0.075 & 0.15 & 0.165 \\
  1&0.02 & 0.04 & 0.09 & 0.15 & \textbf{0.3} & \textbf{0.33} \\
  2&0.04 & 0.08 & 0.12 & \textbf{0.3} & 0.45 & 0.5 \\
  3&0.055 & 0.11 & \textbf{0.3}& 0.45 & 0.51& 0.55\\
  4&0.15 & \textbf{0.3}& 0.46 & 0.5& 0.55 & 0.6
}
\hspace{6ex} 
\scenario{4}{
  0&& 0.025 & 0.045 & 0.06 & 0.08 & 0.15 \\
  1&0.025 & 0.05 & 0.09 & 0.12 & 0.16 & \textbf{0.3} \\
  2&0.08 & 0.16 & \textbf{0.3} & 0.45 & 0.49 & 0.52 \\
  3&0.15 & \textbf{0.3} & 0.46& 0.48 & 0.5& 0.53\\
  4&0.23 & 0.46& 0.48 & 0.5& 0.52 & 0.54
}
\scenario{5}{
  0&& 0.02 & 0.04 & 0.06 & 0.1 & 0.12 \\
  1&0.02 & 0.04 & 0.08 & 0.12 & 0.2 & 0.24 \\
  2&0.05 & 0.1 & 0.14 & 0.18 & 0.22 & \textbf{0.26} \\
  3&0.08 & 0.16 & 0.2& 0.24 & \textbf{0.3}& \textbf{0.34}\\
  4&0.09 & 0.18& \textbf{0.26} & \textbf{0.3}& \textbf{0.34} & 0.36
}
\hspace{6ex}
\scenario{6}{
  0&& 0.02 & 0.065 & 0.125 & 0.165 & 0.195 \\
  1&0.02 & 0.04 & 0.13 & \textbf{0.25} & \textbf{0.33} & 0.39 \\
  2&0.04 & 0.08 & 0.15 & \textbf{0.3} & 0.45 & 0.5 \\
  3&0.055 & 0.11 & 0.21& 0.45 & 0.51& 0.55\\
  4&0.1 & 0.2& \textbf{0.3} & 0.5& 0.55 & 0.65
}
\scenario{7}{
  0&& 0.02 & 0.03 & 0.05 & 0.06 & 0.08 \\
  1&0.02 & 0.04 & 0.06 & 0.1 & 0.12 & 0.16 \\
  2&0.04 & 0.08 & 0.15 & 0.2 & 0.24 & \textbf{0.25} \\
  3&0.055 & 0.11 & 0.2& 0.22 & \textbf{0.26}& \textbf{0.3}\\
  4&0.07 & 0.14& 0.21 & \textbf{0.29}& 0.36 & 0.38
}
\label{tbl:true_toxicity}
\end{table}

 We explore popular OCs used in the literature that are pertinent to the MTDC selection and patient allocation of the MCi3+3, Ci3+3 and POCRML designs.
We define the Probability of Correct Selection (PCS) as the proportion of trials in which the selected dose combination is a true MTDC, and  the Probability of Over or Under Selection (POS or PUS) as the proportion in which the selected dose combination is higher or lower than all true MTDCs, respectively. 
The criteria, Probability of Correct, Over, or Under Allocation (PCA, POA, or PUA) for patient allocation are defined in a similar way. The  OCs for both selection and allocation across the five designs (including two versions of MCi3+3 and  POCRML) are presented in Figures \ref{fig:mtd_selection} and  \ref{fig:patient_allocation}, respectively.

In scenarios 1, 2, 5, and 7, where the true MTDCs occupy the bottom right quadrant, the MCi3+3 design demonstrates slightly lower PCS and comparable PCA than the Ci3+3 design. In scenarios 1 \& 2, the difference of PCS of the MCi3+3 and Ci3+3 designs are within 3\%. Nonetheless, in terms of POA, an important safety criterion, the MCi3+3 design consistently exhibits much smaller POA values than the Ci3+3 design in these scenarios, highlighting the improved safety of the modified approach. The POCRML design underperforms  in scenarios 1, 5, and 7, where the true MTDCs are not positioned in the center of the DC matrix.  Both POCRML-1 and POCRML-5 exhibit performance similar to each other, but with about 30\% reduction compared to the MCi3+3 design in terms of selection accuracy. In scenario 2, the MCi3+3 design demonstrates comparable performance to that of POCRML. Regarding patient allocation, the PCA of POCRML demonstrates a similar trend across these scenarios, yet it exhibits a much lower POA. It seems that the POCRML designs are overly conservative by allocating a large number of patients to doses below the  MTDC in these scenarios.

Scenarios 3, 4, and 6 present a unique case where the true MTDCs span diagonally from the lower left to the upper right of the DC matrix. The MCi3+3 design outperforms the Ci3+3 design in terms of PCS. For instance, in scenario 3, the three-stage MCi3+3 design achieves a PCS of 80.3\%, compared to Ci3+3's 68.7\%. MCi3+3 also features a higher PCA and a lower POA, indicating 
safer patient allocation. In these three scenarios, the POCRML-1 design demonstrates superior PCS and PCA compared to the MCi3+3 design. Notably, in scenario 4, POCRML-1 outperforms the other five designs with a success probability of 82\%, versus 65.4\% for the three-stage MCi3+3. However, it is important to note that the POCRML-1 is the theoretically optimal version of POCRML since it assumes the correct simple order in its model. 
The results of the POCRML-5 and MCi3+3 designs are comparable across these scenarios.


To summarize, the simulation results show that the MCi3+3 design, whether implemented in a two-stage or three-stage format, is comparable to the Ci3+3 design in identifying the MTDC but with much improved safety.
In other words, the new design is able to finding the correct MTDCs with few patients assigned to highly toxic DCs. In addition, the MCi3+3 shows a better performance than the POCRM-5 design in terms of MTDC selection and patients allocation. The optimal version of POCRM, the POCRM-1 design, on the other hand, excels in a few scenarios. However, even with the true simple order. POCRM-1 sometimes can still lead to overly conservative patient allocation and MTDC selection, as shown in scenarios 1, 2, 5, and 7.

In addition to these results, we also conducted simulations based on the scenarios described in \cite{wages2011continual}. The outcomes of these simulations are presented in Appendix \ref{sec:add_sim_res}.

\section{Trial Sample}
\label{sec:trial_sample}
\begin{figure}[!ht]
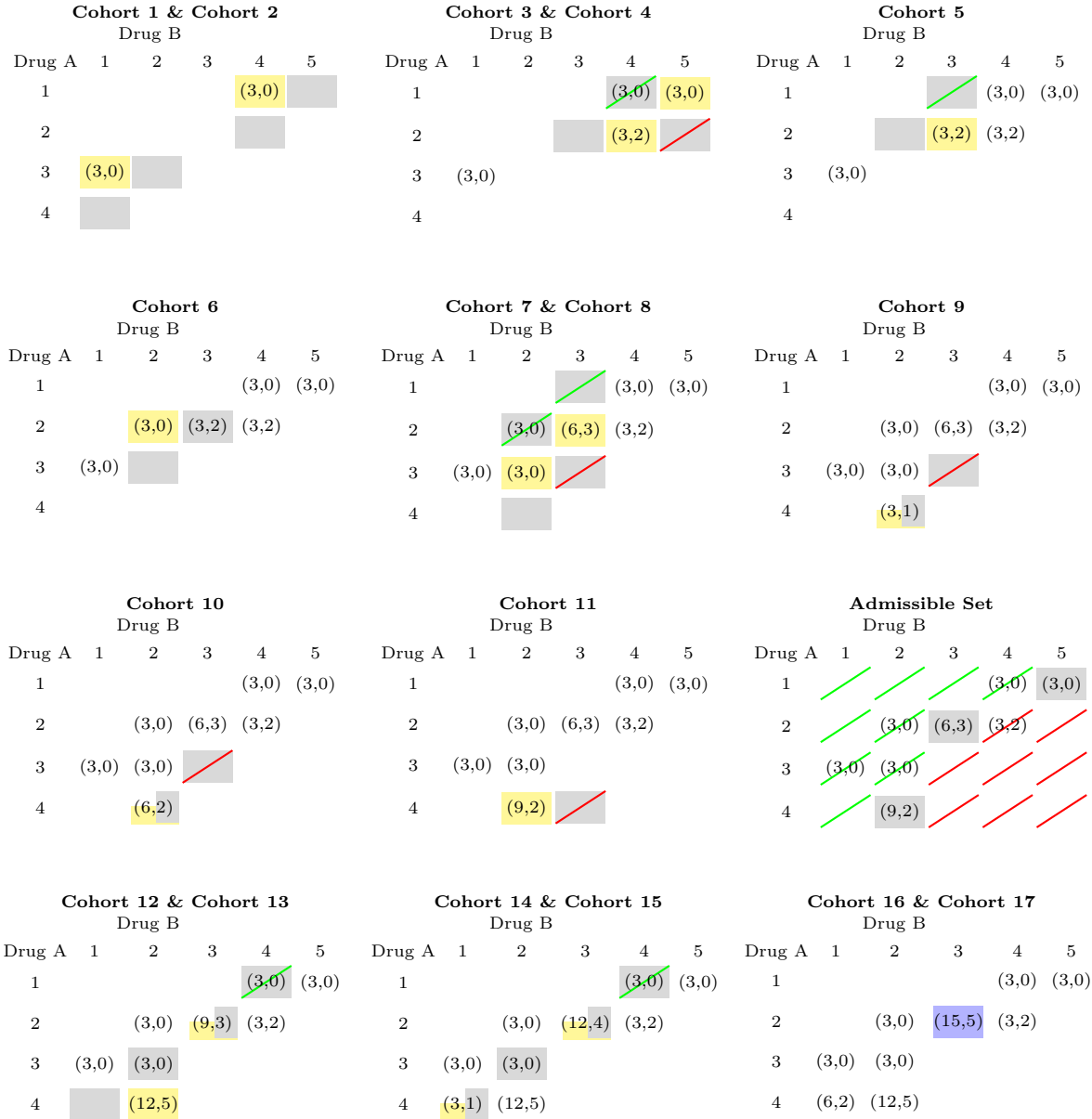

\centering
\setlength{\tabcolsep}{0.1pt}
\renewcommand{\arraystretch}{0.8}

\fontsize{7pt}{8pt}\selectfont

\begin{tabular}{ccc}

\minipage[t]{0.3\textwidth}
\centering
\textbf{Cohort 1 \& Cohort 2}\\
\begin{tabular}{cccccc}
\multicolumn{5}{c}{Drug B} \\
Drug A & 1 & 2 & 3 & 4 & 5 \\
\raisebox{1.5ex}[0pt][0pt]{1} & \customcell{white}{}{} & \customcell{white}{}{} & \customcell{white}{}{} & \customcell{yellow!50}{}{(3,0)} & \customcell{gray!30}{}{} \\
\raisebox{1.5ex}[0pt][0pt]{2} & \customcell{white}{}{} & \customcell{white}{}{} & \customcell{white}{}{} & \customcell{gray!30}{}{} & \customcell{white}{}{} \\
\raisebox{1.5ex}[0pt][0pt]{3} & \customcell{yellow!50}{}{(3,0)}  & \customcell{gray!30}{}{} & \customcell{white}{}{} & \customcell{white}{}{} & \customcell{white}{}{} \\
\raisebox{1.5ex}[0pt][0pt]{4} &\customcell{gray!30}{}{} &  \customcell{white}{}{}&  \customcell{white}{}{}& \customcell{white}{}{}& \customcell{white}{}{}\\
\end{tabular}
\endminipage
&

% ...
\minipage[t]{0.3\textwidth}
\centering
\textbf{Cohort 3 \& Cohort 4} \\
\begin{tabular}{cccccc}
\multicolumn{5}{c}{Drug B} \\
Drug A & 1 & 2 & 3 & 4 & 5 \\
\raisebox{1.5ex}[0pt][0pt]{1} & \customcell{white}{}{} & \customcell{white}{}{} & \customcell{white}{}{} & \customcell{gray!30}{green}{(3,0)} & \customcell{yellow!50}{}{(3,0)} \\
\raisebox{1.5ex}[0pt][0pt]{2} & \customcell{white}{}{} & \customcell{white}{}{} & \customcell{gray!30}{}{} & \customcell{yellow!50}{}{(3,2)} & \customcell{gray!30}{red}{} \\
\raisebox{1.5ex}[0pt][0pt]{3} & \customcell{white}{}{(3,0)}  & \customcell{white}{}{} & \customcell{white}{}{} & \customcell{white}{}{} & \customcell{white}{}{} \\
\raisebox{1.5ex}[0pt][0pt]{4} &\customcell{white}{}{} &  \customcell{white}{}{}&  \customcell{white}{}{}& \customcell{white}{}{}& \customcell{white}{}{}\\
\end{tabular}
\endminipage
&

\minipage[t]{0.3\textwidth}
\centering
\textbf{Cohort 5} \\
\begin{tabular}{cccccc}
\multicolumn{5}{c}{Drug B} \\
Drug A & 1 & 2 & 3 & 4 & 5 \\
\raisebox{1.5ex}[0pt][0pt]{1} & \customcell{white}{}{} & \customcell{white}{}{} & \customcell{gray!30}{green}{} & \customcell{white}{}{(3,0)} & \customcell{white}{}{(3,0)} \\
\raisebox{1.5ex}[0pt][0pt]{2} & \customcell{white}{}{} & \customcell{gray!30}{}{} & \customcell{yellow!50}{}{(3,2)} & \customcell{white}{}{(3,2)} & \customcell{white}{}{} \\
\raisebox{1.5ex}[0pt][0pt]{3} & \customcell{white}{}{(3,0)}  & \customcell{white}{}{} & \customcell{white}{}{} & \customcell{white}{}{} & \customcell{white}{}{} \\
\raisebox{1.5ex}[0pt][0pt]{4} &\customcell{white}{}{} &  \customcell{white}{}{}&  \customcell{white}{}{}& \customcell{white}{}{}& \customcell{white}{}{}\\
\end{tabular}
\endminipage
\end{tabular}

\vspace{20pt}

\begin{tabular}{ccc}
\minipage[t]{0.3\textwidth}
\centering
\textbf{Cohort 6} \\
\begin{tabular}{cccccc}
\multicolumn{5}{c}{Drug B} \\
Drug A & 1 & 2 & 3 & 4 & 5 \\
\raisebox{1.5ex}[0pt][0pt]{1} & \customcell{white}{}{} & \customcell{white}{}{} & \customcell{white}{}{} & \customcell{white}{}{(3,0)} & \customcell{white}{}{(3,0)} \\
\raisebox{1.5ex}[0pt][0pt]{2} & \customcell{white}{}{} & \customcell{yellow!50}{}{(3,0)} & \customcell{gray!30}{}{(3,2)} & \customcell{white}{}{(3,2)} & \customcell{white}{}{} \\
\raisebox{1.5ex}[0pt][0pt]{3} & \customcell{white}{}{(3,0)}  & \customcell{gray!30}{}{} & \customcell{white}{}{} & \customcell{white}{}{} & \customcell{white}{}{} \\
\raisebox{1.5ex}[0pt][0pt]{4} &\customcell{white}{}{} &  \customcell{white}{}{}&  \customcell{white}{}{}& \customcell{white}{}{}& \customcell{white}{}{}\\
\end{tabular}
\endminipage
&

% ...
\minipage[t]{0.3\textwidth}
\centering
\textbf{Cohort 7 \& Cohort 8} \\
\begin{tabular}{cccccc}
\multicolumn{5}{c}{Drug B} \\
Drug A & 1 & 2 & 3 & 4 & 5 \\
\raisebox{1.5ex}[0pt][0pt]{1} & \customcell{white}{}{} & \customcell{white}{}{} & \customcell{gray!30}{green}{} & \customcell{white}{}{(3,0)} & \customcell{white}{}{(3,0)} \\
\raisebox{1.5ex}[0pt][0pt]{2} & \customcell{white}{}{} & \customcell{gray!30}{green}{(3,0)} & \customcell{yellow!50}{}{(6,3)} & \customcell{white}{}{(3,2)} & \customcell{white}{}{} \\
\raisebox{1.5ex}[0pt][0pt]{3} & \customcell{white}{}{(3,0)}  & \customcell{yellow!50}{}{(3,0)} & \customcell{gray!30}{red}{} & \customcell{white}{}{} & \customcell{white}{}{} \\
\raisebox{1.5ex}[0pt][0pt]{4} &\customcell{white}{}{} &  \customcell{gray!30}{}{}&  \customcell{white}{}{}& \customcell{white}{}{}& \customcell{white}{}{}\\
\end{tabular}
\endminipage
&
\minipage[t]{0.3\textwidth}
\centering
\textbf{Cohort 9} \\
\begin{tabular}{cccccc}
\multicolumn{5}{c}{Drug B} \\
Drug A & 1 & 2 & 3 & 4 & 5 \\
\raisebox{1.5ex}[0pt][0pt]{1} & \customcell{white}{}{} & \customcell{white}{}{} & \customcell{white}{}{} & \customcell{white}{white}{(3,0)} & \customcell{white}{}{(3,0)} \\
\raisebox{1.5ex}[0pt][0pt]{2} & \customcell{white}{}{} & \customcell{white}{}{(3,0)} & \customcell{white}{}{(6,3)} & \customcell{white}{}{(3,2)} & \customcell{white}{}{} \\
\raisebox{1.5ex}[0pt][0pt]{3} & \customcell{white}{}{(3,0)}  & \customcell{white}{}{(3,0)} & \customcell{gray!30}{red}{} & \customcell{white}{}{} & \customcell{white}{}{} \\
\raisebox{1.5ex}[0pt][0pt]{4} &\customcell{white}{}{} &  \customcell{halfcolor}{}{(3,1)}&  \customcell{white}{}{}& \customcell{white}{}{}& \customcell{white}{}{}\\
\end{tabular}
\endminipage

\end{tabular}

\vspace{20pt}

\begin{tabular}{ccc}
\minipage[t]{0.3\textwidth}
\centering
\textbf{Cohort 10} \\
\begin{tabular}{cccccc}
\multicolumn{5}{c}{Drug B} \\
Drug A & 1 & 2 & 3 & 4 & 5 \\
\raisebox{1.5ex}[0pt][0pt]{1} & \customcell{white}{}{} & \customcell{white}{}{} & \customcell{white}{}{} & \customcell{white}{white}{(3,0)} & \customcell{white}{}{(3,0)} \\
\raisebox{1.5ex}[0pt][0pt]{2} & \customcell{white}{}{} & \customcell{white}{}{(3,0)} & \customcell{white}{}{(6,3)} & \customcell{white}{}{(3,2)} & \customcell{white}{}{} \\
\raisebox{1.5ex}[0pt][0pt]{3} & \customcell{white}{}{(3,0)}  & \customcell{white}{}{(3,0)} & \customcell{gray!30}{red}{} & \customcell{white}{}{} & \customcell{white}{}{} \\
\raisebox{1.5ex}[0pt][0pt]{4} &\customcell{white}{}{} &  \customcell{halfcolor}{}{(6,2)}&  \customcell{white}{}{}& \customcell{white}{}{}& \customcell{white}{}{}\\
\end{tabular}
\endminipage
&

% ...
\minipage[t]{0.3\textwidth}
\centering
\textbf{Cohort 11} \\
\begin{tabular}{cccccc}
\multicolumn{5}{c}{Drug B} \\
Drug A & 1 & 2 & 3 & 4 & 5 \\
\raisebox{1.5ex}[0pt][0pt]{1} & \customcell{white}{}{} & \customcell{white}{}{} & \customcell{white}{}{} & \customcell{white}{white}{(3,0)} & \customcell{white}{}{(3,0)} \\
\raisebox{1.5ex}[0pt][0pt]{2} & \customcell{white}{}{} & \customcell{white}{}{(3,0)} & \customcell{white}{}{(6,3)} & \customcell{white}{}{(3,2)} & \customcell{white}{}{} \\
\raisebox{1.5ex}[0pt][0pt]{3} & \customcell{white}{}{(3,0)}  & \customcell{white}{}{(3,0)} & \customcell{white}{}{} & \customcell{white}{}{} & \customcell{white}{}{} \\
\raisebox{1.5ex}[0pt][0pt]{4} &\customcell{white}{}{} &  \customcell{yellow!50}{}{(9,2)}&  \customcell{gray!30}{red}{}& \customcell{white}{}{}& \customcell{white}{}{}\\
\end{tabular}
\endminipage
&
\minipage[t]{0.3\textwidth}
\centering
\textbf{Admissible Set} \\
\begin{tabular}{cccccc}
\multicolumn{5}{c}{Drug B} \\
Drug A & 1 & 2 & 3 & 4 & 5 \\
\raisebox{1.5ex}[0pt][0pt]{1} & \customcell{white}{green}{} & \customcell{white}{green}{} & \customcell{white}{green}{} & \customcell{white}{green}{(3,0)} & \customcell{gray!30}{}{(3,0)} \\
\raisebox{1.5ex}[0pt][0pt]{2} & \customcell{white}{green}{} & \customcell{white}{green}{(3,0)} & \customcell{gray!30}{}{(6,3)} & \customcell{white}{red}{(3,2)} & \customcell{white}{red}{} \\
\raisebox{1.5ex}[0pt][0pt]{3} & \customcell{white}{green}{(3,0)}  & \customcell{white}{green}{(3,0)} & \customcell{white}{red}{} & \customcell{white}{red}{} & \customcell{white}{red}{} \\
\raisebox{1.5ex}[0pt][0pt]{4} &\customcell{white}{green}{} &  \customcell{gray!30}{}{(9,2)}&  \customcell{white}{red}{}& \customcell{white}{red}{}& \customcell{white}{red}{}\\
\end{tabular}
\endminipage

\end{tabular}

\vspace{20pt}

\begin{tabular}{ccc}
\minipage[t]{0.3\textwidth}
\centering
\textbf{Cohort 12 \& Cohort 13} \\
\begin{tabular}{cccccc}
\multicolumn{5}{c}{Drug B} \\
Drug A & 1 & 2 & 3 & 4 & 5 \\
\raisebox{1.5ex}[0pt][0pt]{1} & \customcell{white}{}{} & \customcell{white}{}{} & \customcell{white}{}{} & \customcell{gray!30}{green}{(3,0)} & \customcell{white}{}{(3,0)} \\
\raisebox{1.5ex}[0pt][0pt]{2} & \customcell{white}{}{} & \customcell{white}{}{(3,0)} & \customcell{halfcolor}{}{(9,3)} & \customcell{white}{}{(3,2)} & \customcell{white}{}{} \\
\raisebox{1.5ex}[0pt][0pt]{3} & \customcell{white}{}{(3,0)}  & \customcell{gray!30}{}{(3,0)} & \customcell{white}{}{} & \customcell{white}{}{} & \customcell{white}{}{} \\
\raisebox{1.5ex}[0pt][0pt]{4} &\customcell{gray!30}{}{} &  \customcell{yellow!50}{}{(12,5)}&  \customcell{white}{}{}& \customcell{white}{}{}& \customcell{white}{}{}\\
\end{tabular}
\endminipage
&

% ...
\minipage[t]{0.3\textwidth}
\centering
\textbf{Cohort 14 \& Cohort 15} \\
\begin{tabular}{cccccc}
\multicolumn{5}{c}{Drug B} \\
Drug A & 1 & 2 & 3 & 4 & 5 \\
\raisebox{1.5ex}[0pt][0pt]{1} & \customcell{white}{}{} & \customcell{white}{}{} & \customcell{white}{}{} & \customcell{gray!30}{green}{(3,0)} & \customcell{white}{}{(3,0)} \\
\raisebox{1.5ex}[0pt][0pt]{2} & \customcell{white}{}{} & \customcell{white}{}{(3,0)} & \customcell{halfcolor}{}{(12,4)} & \customcell{white}{}{(3,2)} & \customcell{white}{}{} \\
\raisebox{1.5ex}[0pt][0pt]{3} & \customcell{white}{}{(3,0)}  & \customcell{gray!30}{}{(3,0)} & \customcell{white}{}{} & \customcell{white}{}{} & \customcell{white}{}{} \\
\raisebox{1.5ex}[0pt][0pt]{4} &\customcell{halfcolor}{}{(3,1)} &  \customcell{white}{}{(12,5)}&  \customcell{white}{}{}& \customcell{white}{}{}& \customcell{white}{}{}\\
\end{tabular}
\endminipage
&
\minipage[t]{0.3\textwidth}
\centering
\textbf{Cohort 16 \& Cohort 17} \\
\begin{tabular}{cccccc}
\multicolumn{5}{c}{Drug B} \\
Drug A & 1 & 2 & 3 & 4 & 5 \\
\raisebox{1.5ex}[0pt][0pt]{1} & \customcell{white}{}{} & \customcell{white}{}{} & \customcell{white}{}{} & \customcell{white}{}{(3,0)} & \customcell{white}{}{(3,0)} \\
\raisebox{1.5ex}[0pt][0pt]{2} & \customcell{white}{}{} & \customcell{white}{}{(3,0)} & \customcell{blue!30}{}{(15,5)} & \customcell{white}{}{(3,2)} & \customcell{white}{}{} \\
\raisebox{1.5ex}[0pt][0pt]{3} & \customcell{white}{}{(3,0)}  & \customcell{white}{}{(3,0)} & \customcell{white}{}{} & \customcell{white}{}{} & \customcell{white}{}{} \\
\raisebox{1.5ex}[0pt][0pt]{4} &\customcell{white}{}{(6,2)} &  \customcell{white}{}{(12,5)}&  \customcell{white}{}{}& \customcell{white}{}{}& \customcell{white}{}{}\\
\end{tabular}
\endminipage

\end{tabular}
\caption{ A hypothetical trial example demonstrating the MCi3+3 rules in Stage \Rmnum{2}. Yellow boxes are the current DCs; grey ones are candidate DCs for further testing, and boxes with green or red crosses are the DCs not considered for the next cohort of patients. Up to two DCs are tested at each step.
The final MTDC is determined to be $(2,3)$.}
\label{fig:trial_sample}
\end{figure}

We consider a hypothetical trial to demonstrate the decision rules of the MCi3+3 design outlined in Section \ref{sec:mci3p3_design}. In  Figure  \ref{fig:trial_sample} we map out a step-by-step illustration of the MCi3+3 rules based on generated data using the true toxicity probabilities of DC in scenario 3 in Table \ref{tbl:true_toxicity}. We set the target toxicity probability $p_T$ at 0.3 and EI to be between 0.25 and 0.35. We let the  sample size equal 48 for the DC finding phase. We explain each step of Figure \ref{fig:trial_sample} next. There are 12 sub-figures and we use their titles next to label and describe each of them.

\textbf{Cohort 1 \& Cohort 2}: As delineated in Section \ref{sec:single_agent}, the starting DCs for the second stage of MCi3+3 are $(3,1)$ and $(1,4)$, which leads to enrolling two cohorts at these  DCs. Owing to the observed data $(3,0)$ at both DCs, the decisions are $E$ for both DCs. Therefore, according to Rule 3.a in Section \ref{sec:stage2},  DCs $(4,1)$, $(3,2)$, $(2,4)$ and $(1,5)$ are added to the candidate set first, corresponding to the grey boxes in the figure. And DCs $(2,4)$ and $(1,5)$ are selected for the next cohort based on Rule 6.

\textbf{Cohort 3 \& Cohort 4}: Two cohorts of patients are treated at DCs $(2,4)$ and $(1,5)$. Given the observed data $(3,2)$ at DC $(2,4)$ and $(3,0)$ at DC $(1,5)$, the corresponding decisions are $D$ and $E$ respectively. Therefore, DCs $(1,4)$, $(2,3)$ and $(2,5)$ are added to the admissible set first. According to the rule 4.a, DC $(1,4)$ is lower than DC $(1,5)$ with a decision $E$, thus DC $(1,4)$ is eliminated from the candidate set since it is too low. Similarly, DC $(2,5)$ is higher than DC $(2,4)$ with a decision $D$, and therefore DC $(2,5)$ is  eliminated due to Rule 4.b since it is deemed too risky. Consequently, only DC $(2,3)$ remains eligible for future testing.

\textbf{Cohort 5}: One cohort of patients is enrolled at DC $(2,3)$ and the observed outcome is $(3,2)$. The decision for DC $(2,3)$ is $D$, thus DCs $(2,2)$ and $(1,3)$ are included in the candidate set first. DC $(1,3)$ is eliminated since it is lower than DC $(1,4)$ with a decision $E$, according to Rule 4.a, and therefore DC $(2,2)$ is used for further testing.

\textbf{Cohort 6}: The outcome from the new cohort at DC $(2,2)$ is $(3,0)$, and the decision is $E$. Therefore, DCs $(3,2)$ and $(2,3)$ are included in the candidate set. No rules are applied and both DCs are used for further testing.

\textbf{Cohort 7 \& Cohort 8}: According to the new outcomes,
 the decisions for DCs $(2,3)$ and $(3,2)$ are $D$ and $E$, respectively. Therefore,  DCs $(1,3)$, $(2,2)$, $(3,3)$ and $(4,2)$ are included in the candidate set first. However, applying Rule 4.a,  DC $(1,3)$ is eliminated since it is lower than DC $(1,4)$ with a decision $E$ and  DC $(2,2)$ is eliminated as it is lower than DC $(3,2)$ at which the decision is $E$. DC $(3,3)$ is eliminated as it is higher than DC $(2,3)$ with a decision $D$ due to Rule 4.b. Therefore, only DC $(4,2)$ is available for next cohort.

\textbf{Cohort 9}: Based on the outcome at DC $(4,2)$, the decision is $S$. Therefore, DCs $(4,2)$ and $(3,3)$ are included in the candidate set. DC $(3,3)$ is eliminated due to Rule 4.b.  Only DC $(4,2)$ is available for next cohort.

\textbf{Cohort 10}: 
The decision for DC $(4,2)$ is $S$. DC $(4,2)$ and $(3,3)$ are added. DC $(3,3)$ is eliminated again due to Rule 4.b.  Only DC $(4,2)$ is available for next cohort.

\textbf{Cohort 11}: Based on the new outcome, the decision for DC $(4,2)$ is $E$. Therefore, DC $(4,3)$ is included in the candidate set first. However, DC $(4,3)$ is higher than DC $(2,3)$ with a decision $D$. According to Rule 4.b, DC $(4,3)$ is eliminated, and there are no DCs in the candidate set. We then resort to Rule 5.b to identify admissible DCs, which are highlighted in the next sub-figure.

\textbf{Admissible Set}: From Cohort 11, after applying Rule 5.b, the admissible set includes  DCs $(1,5)$, $(2,3)$ and $(4,2)$, among which DCs $(2,3)$ and $(4,2)$ are selected based on their utilities. Therefore, two new cohorts are assigned to the two DCs.

\textbf{Cohort 12 \& Cohort 13}: Based on the outcomes from the new cohorts, the decision assigned to DC $(4,2)$ is $D$, while DC $(2,3)$ receives an $S$. Therefore, DCs $(1,4)$, $(2,3)$, $(3,2)$ and $(4,1)$ are included in the candidate set first. DC $(1,4)$ is subsequently eliminated due to its inferiority to DC $(1,5)$, which was given a decision of $E$. Based on their utilities, DCs $(4,1)$ and $(2,3)$ are then selected. 

\textbf{Cohort 14 \& Cohort 15}: Given the outcome of the new cohorts, the decisions for DCs $(4,1)$ and $(2,3)$ are both $S$. Therefore, DCs $(1,4)$, $(2,3)$, $(3,2)$  and $(4,1)$ are included in the candidate set. However, based on Rule 4.a, DC $(1,4)$ is removed from consideration as it is deemed inferior to DC $(1,5)$, which was previously assigned a $E$ decision. Consequently, DCs $(4,1)$ and $(2,3)$ are chosen based on their utilities.

\textbf{Cohort 16 \& Cohort 17}: Two more cohorts are enrolled at DCs $(4,1)$ and $(2,3)$. And the final MTDC is estimated to be $(2,3)$.

\section{Trial Sample with POCRML}


We also compare the specific decisions made by POCRML using the decision function (pocrm.imp) provided in the R package ``pocrm" based on the same hypothetical trial example in Section \ref{sec:trial_sample}.
We employ the same notation and settings as Section \ref{sec:simulation} for the POCRML designs. 
Here we highlight the dosing decisions based on the observed trial data for each cohort. We use purple box \fcolorbox{purple!50}{purple!50}{\textcolor{purple!50}{test}} to denote decision of MCi3+3, olive box \fcolorbox{olive!50}{olive!50}{\textcolor{olive!50}{test}} 
 of POCRML-5, and cyan box \fcolorbox{cyan!50}{cyan!50}{\textcolor{cyan!50}{test}} of POCRML-1, respectively. 
\begin{figure}[!ht]
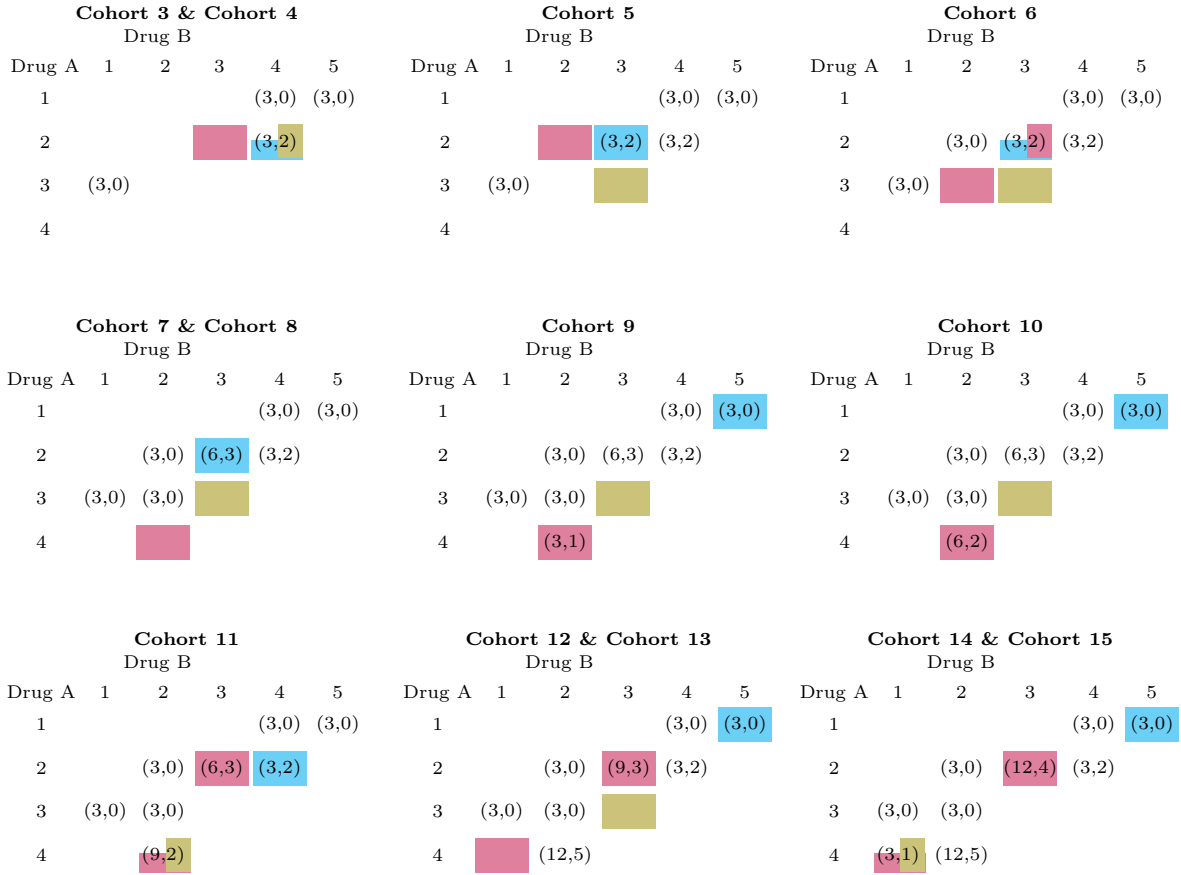

\centering
\setlength{\tabcolsep}{0.1pt}
\renewcommand{\arraystretch}{0.8}

\fontsize{7pt}{8pt}\selectfont

\begin{tabular}{ccc}

% ...
\minipage[t]{0.3\textwidth}
\centering
\textbf{Cohort 3 \& Cohort 4} \\
\begin{tabular}{cccccc}
\multicolumn{5}{c}{Drug B} \\
Drug A & 1 & 2 & 3 & 4 & 5 \\
\raisebox{1.5ex}[0pt][0pt]{1} & \customcell{white}{}{} & \customcell{white}{}{} & \customcell{white}{}{} & \customcell{white}{}{(3,0)} & \customcell{white}{}{(3,0)} \\
\raisebox{1.5ex}[0pt][0pt]{2} & \customcell{white}{}{} & \customcell{white}{}{} & \customcell{purple!50}{}{} & \customcell{halfco}{}{(3,2)} & \customcell{white}{}{} \\
\raisebox{1.5ex}[0pt][0pt]{3} & \customcell{white}{}{(3,0)}  & \customcell{white}{}{} & \customcell{white}{}{} & \customcell{white}{}{} & \customcell{white}{}{} \\
\raisebox{1.5ex}[0pt][0pt]{4} &\customcell{white}{}{} &  \customcell{white}{}{}&  \customcell{white}{}{}& \customcell{white}{}{}& \customcell{white}{}{}\\
\end{tabular}
\endminipage
&

\minipage[t]{0.3\textwidth}
\centering
\textbf{Cohort 5} \\
\begin{tabular}{cccccc}
\multicolumn{5}{c}{Drug B} \\
Drug A & 1 & 2 & 3 & 4 & 5 \\
\raisebox{1.5ex}[0pt][0pt]{1} & \customcell{white}{}{} & \customcell{white}{}{} & \customcell{white}{}{} & \customcell{white}{}{(3,0)} & \customcell{white}{}{(3,0)} \\
\raisebox{1.5ex}[0pt][0pt]{2} & \customcell{white}{}{} & \customcell{purple!50}{}{} & \customcell{cyan!50}{}{(3,2)} & \customcell{white}{}{(3,2)} & \customcell{white}{}{} \\
\raisebox{1.5ex}[0pt][0pt]{3} & \customcell{white}{}{(3,0)}  & \customcell{white}{}{} & \customcell{olive!50}{}{} & \customcell{white}{}{} & \customcell{white}{}{} \\
\raisebox{1.5ex}[0pt][0pt]{4} &\customcell{white}{}{} &  \customcell{white}{}{}&  \customcell{white}{}{}& \customcell{white}{}{}& \customcell{white}{}{}\\
\end{tabular}
\endminipage
&
\minipage[t]{0.3\textwidth}
\centering
\textbf{Cohort 6} \\
\begin{tabular}{cccccc}
\multicolumn{5}{c}{Drug B} \\
Drug A & 1 & 2 & 3 & 4 & 5 \\
\raisebox{1.5ex}[0pt][0pt]{1} & \customcell{white}{}{} & \customcell{white}{}{} & \customcell{white}{}{} & \customcell{white}{}{(3,0)} & \customcell{white}{}{(3,0)} \\
\raisebox{1.5ex}[0pt][0pt]{2} & \customcell{white}{}{} & \customcell{white}{}{(3,0)} & \customcell{halfc}{}{(3,2)} & \customcell{white}{}{(3,2)} & \customcell{white}{}{} \\
\raisebox{1.5ex}[0pt][0pt]{3} & \customcell{white}{}{(3,0)}  & \customcell{purple!50}{}{} & \customcell{olive!50}{}{} & \customcell{white}{}{} & \customcell{white}{}{} \\
\raisebox{1.5ex}[0pt][0pt]{4} &\customcell{white}{}{} &  \customcell{white}{}{}&  \customcell{white}{}{}& \customcell{white}{}{}& \customcell{white}{}{}\\
\end{tabular}
\endminipage

\end{tabular}

\vspace{20pt}

\begin{tabular}{ccc}

% ...
\minipage[t]{0.3\textwidth}
\centering
\textbf{Cohort 7 \& Cohort 8} \\
\begin{tabular}{cccccc}
\multicolumn{5}{c}{Drug B} \\
Drug A & 1 & 2 & 3 & 4 & 5 \\
\raisebox{1.5ex}[0pt][0pt]{1} & \customcell{white}{}{} & \customcell{white}{}{} & \customcell{white}{}{} & \customcell{white}{}{(3,0)} & \customcell{white}{}{(3,0)} \\
\raisebox{1.5ex}[0pt][0pt]{2} & \customcell{white}{}{} & \customcell{white}{}{(3,0)} & \customcell{cyan!50}{}{(6,3)} & \customcell{white}{}{(3,2)} & \customcell{white}{}{} \\
\raisebox{1.5ex}[0pt][0pt]{3} & \customcell{white}{}{(3,0)}  & \customcell{white}{}{(3,0)} & \customcell{olive!50}{}{} & \customcell{white}{}{} & \customcell{white}{}{} \\
\raisebox{1.5ex}[0pt][0pt]{4} &\customcell{white}{}{} &  \customcell{purple!50}{}{}&  \customcell{white}{}{}& \customcell{white}{}{}& \customcell{white}{}{}\\
\end{tabular}
\endminipage
&
\minipage[t]{0.3\textwidth}
\centering
\textbf{Cohort 9} \\
\begin{tabular}{cccccc}
\multicolumn{5}{c}{Drug B} \\
Drug A & 1 & 2 & 3 & 4 & 5 \\
\raisebox{1.5ex}[0pt][0pt]{1} & \customcell{white}{}{} & \customcell{white}{}{} & \customcell{white}{}{} & \customcell{white}{}{(3,0)} & \customcell{cyan!50}{}{(3,0)} \\
\raisebox{1.5ex}[0pt][0pt]{2} & \customcell{white}{}{} & \customcell{white}{}{(3,0)} & \customcell{white}{}{(6,3)} & \customcell{white}{}{(3,2)} & \customcell{white}{}{} \\
\raisebox{1.5ex}[0pt][0pt]{3} & \customcell{white}{}{(3,0)}  & \customcell{white}{}{(3,0)} & \customcell{olive!50}{}{} & \customcell{white}{}{} & \customcell{white}{}{} \\
\raisebox{1.5ex}[0pt][0pt]{4} &\customcell{white}{}{} &  \customcell{purple!50}{}{(3,1)}&  \customcell{white}{}{}& \customcell{white}{}{}& \customcell{white}{}{}\\
\end{tabular}
\endminipage
&
\minipage[t]{0.3\textwidth}
\centering
\textbf{Cohort 10} \\
\begin{tabular}{cccccc}
\multicolumn{5}{c}{Drug B} \\
Drug A & 1 & 2 & 3 & 4 & 5 \\
\raisebox{1.5ex}[0pt][0pt]{1} & \customcell{white}{}{} & \customcell{white}{}{} & \customcell{white}{}{} & \customcell{white}{}{(3,0)} & \customcell{cyan!50}{}{(3,0)} \\
\raisebox{1.5ex}[0pt][0pt]{2} & \customcell{white}{}{} & \customcell{white}{}{(3,0)} & \customcell{white}{}{(6,3)} & \customcell{white}{}{(3,2)} & \customcell{white}{}{} \\
\raisebox{1.5ex}[0pt][0pt]{3} & \customcell{white}{}{(3,0)}  & \customcell{white}{}{(3,0)} & \customcell{olive!50}{}{} & \customcell{white}{}{} & \customcell{white}{}{} \\
\raisebox{1.5ex}[0pt][0pt]{4} &\customcell{white}{}{} &  \customcell{purple!50}{}{(6,2)}&  \customcell{white}{}{}& \customcell{white}{}{}& \customcell{white}{}{}\\
\end{tabular}
\endminipage

\end{tabular}

\vspace{20pt}

\begin{tabular}{ccc}

% ...
\minipage[t]{0.3\textwidth}
\centering
\textbf{Cohort 11} \\
\begin{tabular}{cccccc}
\multicolumn{5}{c}{Drug B} \\
Drug A & 1 & 2 & 3 & 4 & 5 \\
\raisebox{1.5ex}[0pt][0pt]{1} & \customcell{white}{}{} & \customcell{white}{}{} & \customcell{white}{}{} & \customcell{white}{}{(3,0)} & \customcell{white}{}{(3,0)} \\
\raisebox{1.5ex}[0pt][0pt]{2} & \customcell{white}{}{} & \customcell{white}{}{(3,0)} & \customcell{purple!50}{}{(6,3)} & \customcell{cyan!50}{}{(3,2)} & \customcell{white}{}{} \\
\raisebox{1.5ex}[0pt][0pt]{3} & \customcell{white}{}{(3,0)}  & \customcell{white}{}{(3,0)} & \customcell{white}{}{} & \customcell{white}{}{} & \customcell{white}{}{} \\
\raisebox{1.5ex}[0pt][0pt]{4} &\customcell{white}{}{} &  \customcell{halfcon}{}{(9,2)}&  \customcell{white}{}{}& \customcell{white}{}{}& \customcell{white}{}{}\\
\end{tabular}
\endminipage
&
\minipage[t]{0.3\textwidth}
\centering
\textbf{Cohort 12 \& Cohort 13} \\
\begin{tabular}{cccccc}
\multicolumn{5}{c}{Drug B} \\
Drug A & 1 & 2 & 3 & 4 & 5 \\
\raisebox{1.5ex}[0pt][0pt]{1} & \customcell{white}{}{} & \customcell{white}{}{} & \customcell{white}{}{} & \customcell{white}{}{(3,0)} & \customcell{cyan!50}{}{(3,0)} \\
\raisebox{1.5ex}[0pt][0pt]{2} & \customcell{white}{}{} & \customcell{white}{}{(3,0)} & \customcell{purple!50}{}{(9,3)} & \customcell{white}{}{(3,2)} & \customcell{white}{}{} \\
\raisebox{1.5ex}[0pt][0pt]{3} & \customcell{white}{}{(3,0)}  & \customcell{white}{}{(3,0)} & \customcell{olive!50}{}{} & \customcell{white}{}{} & \customcell{white}{}{} \\
\raisebox{1.5ex}[0pt][0pt]{4} &\customcell{purple!50}{}{} &  \customcell{white}{}{(12,5)}&  \customcell{white}{}{}& \customcell{white}{}{}& \customcell{white}{}{}\\
\end{tabular}
\endminipage
&
\minipage[t]{0.3\textwidth}
\centering
\textbf{Cohort 14 \& Cohort 15} \\
\begin{tabular}{cccccc}
\multicolumn{5}{c}{Drug B} \\
Drug A & 1 & 2 & 3 & 4 & 5 \\
\raisebox{1.5ex}[0pt][0pt]{1} & \customcell{white}{}{} & \customcell{white}{}{} & \customcell{white}{}{} & \customcell{white}{}{(3,0)} & \customcell{cyan!50}{}{(3,0)} \\
\raisebox{1.5ex}[0pt][0pt]{2} & \customcell{white}{}{} & \customcell{white}{}{(3,0)} & \customcell{purple!50}{}{(12,4)} & \customcell{white}{}{(3,2)} & \customcell{white}{}{} \\
\raisebox{1.5ex}[0pt][0pt]{3} & \customcell{white}{}{(3,0)}  & \customcell{white}{}{(3,0)} & \customcell{white}{}{} & \customcell{white}{}{} & \customcell{white}{}{} \\
\raisebox{1.5ex}[0pt][0pt]{4} &\customcell{halfcon}{}{(3,1)} &  \customcell{white}{}{(12,5)}&  \customcell{white}{}{}& \customcell{white}{}{}& \customcell{white}{}{}\\
\end{tabular}
\endminipage
\end{tabular}

\caption{ A comparison between MCi3+3 and POCRML based on trial data. Here a purple box \fcolorbox{purple!50}{purple!50}{\textcolor{purple!50}{test}}  denotes decision of MCi3+3, an olive box \fcolorbox{olive!50}{olive!50}{\textcolor{olive!50}{test}} 
 of POCRML-5, and a cyan box \fcolorbox{cyan!50}{cyan!50}{\textcolor{cyan!50}{test}} of POCRML-1, respectively.}
\label{fig:trial_sample_pocrm}
\end{figure}

 The decisions from POCRML-5 for Cohort 5, 6, 7$\&$ 8, 9 and 10 are risky, escalating dose even in the presence of excessive number of DLTs at a lower DC.  
POCRML-1 performs better although it it also gives  a risky decision in Cohort 11, where it  allocates patients DC (2,4) where three out of six patients at DC (2,3)  experienced DLT. 
\jj

\section{Discussion}
\label{sec:discussion}



Existing Phase I designs for dual-agent trials do not support parallel allocation and may under-utilize historical single-phase data. In response, we introduce the MCi3+3 design, a versatile framework accommodating a two or three-stage approach. It commences with single-agent dose finding to determine the starting dose for the combination phase. The subsequent stage employs a rule-based design that facilitates parallel testing of up to two doses, and, should the trial advance to the third stage, a model-based approach is adopted to steer patient allocation. Upon trial completion, data from both the single-agent and combination phases are integrated to identify the MTDC. Additionally, we offer the flexibility to select multiple MTDCs. For example, it is conceivable that various dose combinations may fall within the Equivalence Interval (EI), which is also illustrated in Table \ref{tbl:true_toxicity}. Therefore, if the option for multiple MTDCs is employed, all the DCs with posterior toxicity values falling within the EIs will be selected as MTDCs.

In our proposed MCi3+3 design, while the single-agent dose-finding stage is designed for parallel patient enrollment, this is not a strict requirement. This means that the trial can commence at different time points. Historical data can still be utilized, as the data from the single-agent phase are crucial for determining the starting dose of the second stage and for modeling in the third stage. Additionally, while we maintain a consistent target toxicity level across both the single-agent and combination phases, this parameter can be adjusted to enhance flexibility in the trial design. Furthermore, although the starting doses for the second stage are typically derived from the first stage, they can also be specified by practitioners. This decision might be based on prior knowledge of the drug's toxicity profile or lower doses, such as $(1,1)$, if required by regulatory authorities.

In the MCi3+3 design, the primary decision-making is guided by the i3+3 design. However, other Phase I designs  can also be considered. This flexibility allows for a more tailored approach to dose determination, accommodating various clinical and regulatory considerations.

\section*{Acknowledgement}
\label{sec6}
The authors thank Brendan Weiss, Lili Zhu, Xuezhou Mao, and Sarah Sloan from Moderna, Inc. for their discussions and suggestions on the method proposed in this manuscript.

\section*{Disclaimer}
\label{sec7}

The opinions expressed in this paper are solely those of the authors and not those of their affiliations. The authors’ affiliations do not guarantee the accuracy or reliability of the information provided herein. \\
Qiqi Deng is employee of and hold stock and options in Moderna, Inc.

\bibliographystyle{apalike}
\bibliography{ci3+3}

\clearpage\newpage

\begin{appendices}
\appendixpage
\setcounter{table}{0}
\renewcommand{\thetable}{A.\arabic{table}}
\setcounter{equation}{0}
\renewcommand{\theequation}{A.\arabic{equation}}
\setcounter{figure}{0}
\renewcommand{\thefigure}{A.\arabic{figure}}

\section{Partial Order of POCRML}\label{sec:partial_order}
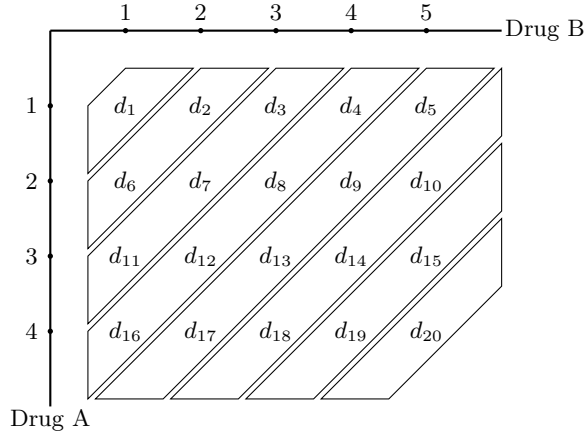
\begin{figure}[h!]
\centering
\begin{tikzpicture}[scale=1, every node/.style={font=\small, inner sep=1pt}]
    \draw[thick, -] (0,0) -- (6,0) node[right] {Drug B};
    \draw[thick, -] (0,0) -- (0,-5) node[below] {Drug A};

    \foreach \x in {1,2,3,4,5} {
        \fill (\x, 0) circle (1pt); 
        \node at (\x, 0.25) {$\x$}; 
    }
    \foreach \y in {1,2,3,4} {
        \fill (0, -\y) circle (1pt); 
        \node at (-0.25, -\y) {$\y$}; 
    }

    \foreach \x in {1,...,5}
    \foreach \y in {1,...,4}
    {
        \pgfmathtruncatemacro{\label}{(\y-1)*5+\x}
        \node at (\x,-\y) {$d_{\label}$};
    }

    \draw (1,-0.5) -- (1.9,-0.5) -- (0.5,-1.9) -- (0.5,-1) -- cycle;
    \draw (2,-0.5) -- (2.9,-0.5) -- (0.5,-2.9) -- (0.5,-2) -- cycle;
    \draw (3,-0.5) -- (3.9,-0.5) -- (0.5,-3.9) -- (0.5,-3) -- cycle;
    \draw (4,-0.5) -- (4.9,-0.5) -- (0.5,-4.9) -- (0.5,-4) -- cycle;
    \draw (5,-0.5) -- (5.9,-0.5) -- (1.5,-4.9) -- (0.6,-4.9) -- cycle;
    \draw (6,-0.5) -- (6,-1.4) -- (2.5,-4.9) -- (1.6,-4.9) -- cycle;
    \draw (6,-1.5) -- (6,-2.4) -- (3.5,-4.9) -- (2.6,-4.9) -- cycle;
    \draw (6,-2.5) -- (6,-3.4) -- (4.5,-4.9) -- (3.6,-4.9) -- cycle;

\end{tikzpicture}
\caption{An illustration of zoning a drug combination matrix.}
\label{fig:partial_order}
\end{figure}

The POCRM design proposes  zoning the drug combination matrix as illustrated in Figure \ref{fig:partial_order} \citep{wages2011continual}. Each zone consists of one or more DCs and represents a potential range of toxicity levels. From upper left to lower right, the toxicity is assumed to increase across zones while DCs within the same zone are assumed to be unordered. This structure produces a partial order of all the DCs. 






POCRML is based on putting a probability model on the potential simple orders corresponding to the partial order in Figure \ref{fig:partial_order}. 
Following the examples in \citep{wages2011continual} and \citep{wages2011dosefinding}, we propose the following five simple   orders for the simulations in Section \ref{sec:simulation}. In this paper, we assume two settings of POCRML based on the assumed simple orders. Setting one assumes five possible simple orders, the corresponding design labeled as  `POCRML-5'. In particular, denote $m \in \{1,2, 3,4,5\}$ the index of the simple order. Assume

\noindent
$m = 1$: $d_1 \rightarrow d_2 \rightarrow d_6 \rightarrow d_3 \rightarrow d_7 \rightarrow d_{11} \rightarrow d_4 \rightarrow d_8 \rightarrow d_{12} \rightarrow d_{16} \rightarrow d_5 \rightarrow d_9 \rightarrow d_{13} \rightarrow d_{17} \rightarrow d_{10} \rightarrow d_{14} \rightarrow d_{18} \rightarrow d_{15} \rightarrow d_{19} \rightarrow d_{20}$

\noindent
$m = 2$: $d_1 \rightarrow d_6 \rightarrow d_2 \rightarrow d_3 \rightarrow d_{11} \rightarrow d_7 \rightarrow d_4 \rightarrow d_8 \rightarrow d_{16} \rightarrow d_{12} \rightarrow d_9 \rightarrow d_5 \rightarrow d_{17} \rightarrow d_{13} \rightarrow d_{10} \rightarrow d_{18} \rightarrow d_{14} \rightarrow d_{19} \rightarrow d_{15} \rightarrow d_{20}$

\noindent
$m = 3$: $d_1 \rightarrow d_6 \rightarrow d_2 \rightarrow d_7 \rightarrow d_3 \rightarrow d_{11} \rightarrow d_8 \rightarrow d_4 \rightarrow d_{12} \rightarrow d_{16} \rightarrow d_5 \rightarrow d_9 \rightarrow d_{17} \rightarrow d_{13} \rightarrow d_{14} \rightarrow d_{10} \rightarrow d_{18} \rightarrow d_{15} \rightarrow d_{19} \rightarrow d_{20}$

\noindent
$m = 4$: $d_1 \rightarrow d_2 \rightarrow d_6 \rightarrow d_{11} \rightarrow d_3 \rightarrow d_7 \rightarrow d_{16} \rightarrow d_{12} \rightarrow d_4 \rightarrow d_8 \rightarrow d_{13} \rightarrow d_{17} \rightarrow d_5 \rightarrow d_9 \rightarrow d_{18} \rightarrow d_{14} \rightarrow d_{10} \rightarrow d_{19} \rightarrow d_{15} \rightarrow d_{20}$

\noindent
$m = 5$: $d_1 \rightarrow d_2 \rightarrow d_6 \rightarrow d_{11} \rightarrow d_7 \rightarrow d_3 \rightarrow d_{12} \rightarrow d_4 \rightarrow d_8 \rightarrow d_{16} \rightarrow d_{17} \rightarrow d_5 \rightarrow d_9 \rightarrow d_{13} \rightarrow d_{18} \rightarrow d_{10} \rightarrow d_{14} \rightarrow d_{15} \rightarrow d_{19} \rightarrow d_{20}$

Additionally, we consider setting two in which there is one simple order of  POCRML, referred to as `POCRML-1'. We assume the simple order of POCRML-1 is exactly the same as the true order for the simulation scenario. This provides the theoretically optimal performance of POCRML. 

The true simple order for each scenario  and POCRML-1 is as follows,

\noindent
$sc = 1$: $d_1 \rightarrow d_2 \rightarrow d_6 \rightarrow d_3 \rightarrow d_7 \rightarrow d_{11} \rightarrow d_4 \rightarrow d_8 \rightarrow d_5 \rightarrow d_{12} \rightarrow d_{16} \rightarrow d_9 \rightarrow d_{13} \rightarrow d_{17} \rightarrow d_{10} \rightarrow d_{14} \rightarrow d_{18} \rightarrow d_{15} \rightarrow d_{19} \rightarrow d_{20}$

\noindent
$sc = 2$: $d_1 \rightarrow d_6 \rightarrow d_{11} \rightarrow d_{16} \rightarrow d_2 \rightarrow d_7 \rightarrow d_{12} \rightarrow d_{17} \rightarrow d_3 \rightarrow d_8 \rightarrow d_4 \rightarrow d_5 \rightarrow d_9 \rightarrow d_{13} \rightarrow d_{18} \rightarrow d_{14} \rightarrow d_{10} \rightarrow d_{15} \rightarrow d_{19} \rightarrow d_{20}$

\noindent
$sc = 3$: $d_1 \rightarrow d_6 \rightarrow d_2 \rightarrow d_{11} \rightarrow d_7 \rightarrow d_3 \rightarrow d_{16} \rightarrow d_{12} \rightarrow d_4 \rightarrow d_8 \rightarrow d_5 \rightarrow d_9 \rightarrow d_{13} \rightarrow d_{17} \rightarrow d_{18} \rightarrow d_{10} \rightarrow d_{14} \rightarrow d_{19} \rightarrow d_{15} \rightarrow d_{20}$

\noindent
$sc = 4$: $d_1 \rightarrow d_2 \rightarrow d_3 \rightarrow d_6 \rightarrow d_4 \rightarrow d_7 \rightarrow d_{11} \rightarrow d_5 \rightarrow d_8 \rightarrow d_{12} \rightarrow d_{16} \rightarrow d_{13} \rightarrow d_{17} \rightarrow d_9 \rightarrow d_{18} \rightarrow d_{14} \rightarrow d_{10} \rightarrow d_{19} \rightarrow d_{15} \rightarrow d_{20}$

\noindent
$sc = 5$: $d_1 \rightarrow d_2 \rightarrow d_6 \rightarrow d_3 \rightarrow d_7 \rightarrow d_{11} \rightarrow d_8 \rightarrow d_{16} \rightarrow d_{12} \rightarrow d_4 \rightarrow d_9 \rightarrow d_{13} \rightarrow d_5 \rightarrow d_{17} \rightarrow d_{10} \rightarrow d_{18} \rightarrow d_{14} \rightarrow d_{19} \rightarrow d_{15} \rightarrow d_{20}$

\noindent
$sc = 6$: $d_1 \rightarrow d_6 \rightarrow d_{11} \rightarrow d_2 \rightarrow d_7 \rightarrow d_{16} \rightarrow d_{12} \rightarrow d_3 \rightarrow d_8 \rightarrow d_{17} \rightarrow d_4 \rightarrow d_5 \rightarrow d_9 \rightarrow d_{13} \rightarrow d_{10} \rightarrow d_{18} \rightarrow d_{14} \rightarrow d_{15} \rightarrow d_{19} \rightarrow d_{20}$

\noindent
$sc = 7$: $d_1 \rightarrow d_2 \rightarrow d_6 \rightarrow d_3 \rightarrow d_{11} \rightarrow d_4 \rightarrow d_{16} \rightarrow d_7 \rightarrow d_5 \rightarrow d_{12} \rightarrow d_8 \rightarrow d_{17} \rightarrow d_{13} \rightarrow d_9 \rightarrow d_{10} \rightarrow d_{14} \rightarrow d_{18} \rightarrow d_{15} \rightarrow d_{19} \rightarrow d_{20}$

\section{Additional Simulation Results} \label{sec:add_sim_res}

In addition to the results presented in Section \ref{sec:simulation}, we also conducted simulations using the scenarios described in \cite{wages2011continual}, employing the same configuration of 2,000 simulations. We compare MCi3+3 and POCRM in \cite{wages2011continual}.Regarding sample size, the POCRM design uses 60 patients for the combination phase; therefore, we set a total sample size of 78 for the MCi3+3 design, resulting in 57-60 patients in the combination phase, which we consider to be a fair comparison. Since the MCi3+3 design is interval-based, we set $\epsilon_1=0.1$ and $\epsilon_2=0$ for the intervals, based on the scenarios. Also, we set the starting dose of MCi3+3 from DC (1,1) to be compatible with POCRM.

\begin{table}[ht]
\centering
\caption{Comparison of POCRM and three-stage MCi3+3 for matrix orders for 6 toxicity scenarios}
\begin{tabular}{lcccccc}
\toprule
Method & Scenario 1 & Scenario 2 & Scenario 3 & Scenario 4 & Scenario 5 & Scenario 6 \\
\midrule
\multicolumn{7}{c}{\% correct recommendation} \\
\midrule
MCi3+3 & 51.90 & 57.65 & 64.25 & 59.55 & 59.90 & 49.20 \\
POCRM        & 53.0 & 50.0 & 86.0 & 68.0 & 38.0 & 45.0 \\
\midrule
\multicolumn{7}{c}{\% of observed toxicities} \\
\midrule
MCi3+3 & 22.1 & 21.6 & 21.7 & 24.5 & 36.2 & 32.8 \\
POCRM        & 31.1 & 31.1 & 31.3 & 33.4 & 46.0 & 44.0 \\
\midrule
\multicolumn{7}{c}{\% patients treated at MTD} \\
\midrule
MCi3+3 & 26.1 & 31.6 & 36.0 & 32.9 & 43.8 & 33.0 \\
POCRM        & 39.0 & 37.0 & 69.0 & 43.0 & 27.0 & 33.0 \\
\bottomrule
\label{tbl:results_pocrm}
\end{tabular}
\end{table}

The simulation results are presented in Table \ref{tbl:results_pocrm}. Overall, the two designs perform similarly across the six scenarios, with one design over-performing in some scenarios but under-performing in others. 

\jj
\clearpage
\newpage

\end{appendices}

\end{document}